# Optochemically Responsive 2D Nanosheets of a 3D Metal-Organic Framework Material

*Abhijeet K. Chaudhari,*[1] *Ha Jin Kim,*[2] *Intaek Han,*[2] and *Jin-Chong Tan*[1]*

[1]*Department of Engineering Science, University of Oxford, OX1 3PJ, Oxford, United Kingdom*
[2]*Samsung Advanced Institute of Technology (SAIT), Samsung Electronics Co. Ltd., South Korea*

*Correspondence to:* jin-chong.tan@eng.ox.ac.uk

**Abstract:**

Outstanding functional tunability underpinning metal-organic framework (MOF) confers a versatile platform to contrive next-generation chemical sensors, optoelectronics, energy harvesters and converters. We report a rare exemplar of a porous 2D nanosheet material, constructed from an extended 3D MOF structure. We develop a rapid supramolecular self-assembly methodology at ambient conditions, to synthesize readily-exfoliatable MOF nanosheets, functionalized *in situ* by adopting the Guest@MOF (Host) strategy. Nanoscale confinement of light-emitting molecules (as functional guest) inside the MOF pores generates unusual combination of optical, electronic, and chemical properties, arising from the strong host-guest coupling effects. We show highly promising photonics based chemical sensing opened up by the new Guest@MOF composite systems. By harnessing host-guest optochemical interactions of functionalized MOF nanosheets, we have accomplished detection of an extensive range of volatile organic compounds (VOCs) and small molecules important for many practical applications.

Two-dimensional (2D) nanosheets are contemporary materials with exceptional physical and functional properties, derived from a broad class of low-dimensional solids containing atomically thin structures[1], exfoliated 2D frameworks[2], and molecular membranes[3]. Considerable efforts are being devoted to 2D graphene-related materials[4] to yield improved mechanical, electronic and optical modulations, important for device integration and disruptive technologies. On the contrary, far less widespread are thin-layered materials derived from self-assembled supramolecular systems[5], whose weak interlayer non-covalent chemical interactions are van der Waals in nature. Such molecule-based 2D materials may provide significant benefits towards physical and chemical tunability, structural flexibility, and ease of exfoliation[6]. There is recent intensifying interest in an emergent class of 2D nanosheets constructed by *downsizing* 3D metal-organic frameworks (MOFs)[3a,7], which are inorganic-organic (hybrid) structures possessing an enormous physicochemical[8] and structural versatility[9]. Furthermore, the nanoscale porosity of MOFs could function as a vessel to "host" a variety of functional "guest" molecules[10], imparting a unique combination of properties through intimate host-guest interactions[11]. Yet, preparation of functionalized MOF as 2D nanosheets exemplifying *tunable host-guest sensing* response is uncommon in literature, unlike its traditional counterparts[12].

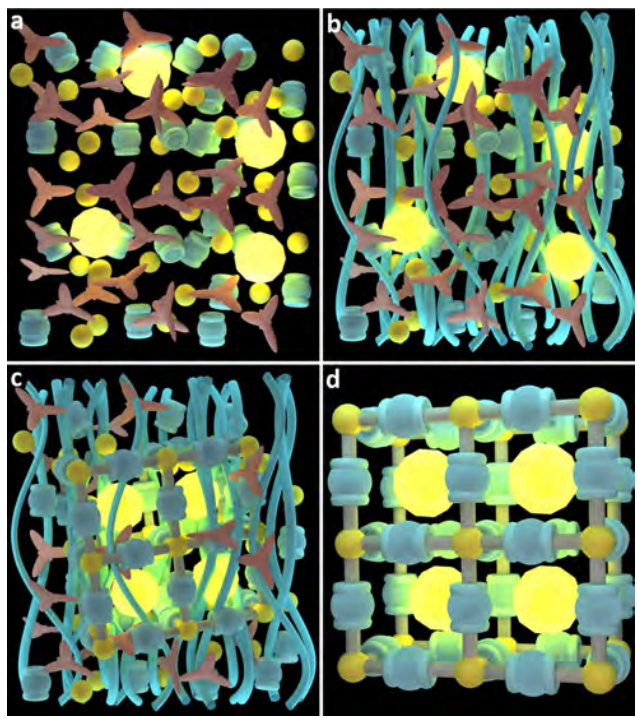

**Scheme 1. Proposed supramolecular synthetic strategy for constructing an optochemically active Guest@MOF (Host) system.** (a) *In situ* assembly began with a one-pot solution mixture of high-concentration reactants, combined with functional guest species (*e.g.* luminescent complexes represented by yellow spheres). (b) This solution mixture produces a fibrous supramolecular assembly, facilitating formation of (c) periodic MOF structures encapsulating functional guest molecules. (d) Disassembly of scaffolding fibers to harvest functionalized Guest@MOF composite systems.

**Supramolecular synthesis and structure of MOF nanosheets**

In this work, we present a simple supramolecular self-assembly strategy to accomplish concomitant 2D nanosheet synthesis and functionalization of a porous MOF system, and demonstrate its efficacy for application as a tunable optochemical sensor. We leverage our recently elucidated supramolecular "high-concentration reactions" (HCR) approach[13], to realize one-pot synthesis of functionalized MOF nanosheets at ambient conditions; the basic concept is illustrated in Scheme 1. Here we describe a representative study employing 1,4-benzenedicarboxylic acid (BDC) as the organic linker, because of its strong propensity to construct an extended chemical network upon coordination with metal centers; for instance, here we utilized divalent $Zn^{2+}$. Triethylamine base ($NEt_3$) featuring a flexible tripodal geometry was used to trigger fast activation (deprotonation) of the BDC linkers[14]. It is striking to see that, a white gel-like fibrous soft matter was immediately obtained at room temperature (Fig. 1a), arising from the high-concentration reaction (HCR) between $Zn^{2+}$ and $BDC^{2-}$, augmented by the $NEt_3^+$ cations. We observed a discernible two-stage material transformation *via* optical microscopy (Fig. 1b): initially witnessing development of highly oriented fibers, prior to formation of a visually *shiny* phase prevalent on the fiber surfaces. The gel fiber diameter was found to be ~1 to 10s μm by scanning electron microscopy (SEM), see Fig. 1c. Intriguingly, SEM revealed those supramolecular fibers are, in fact, constituting densely packed crystalline nanosheets (Fig. 1d), thus confirming the (shiny) faceted appearance detected under optical microscopy (Fig. 1b). To establish the detailed 2D morphologies, we examined nanosheets harvested from the supramolecular gels using transmission electron microscopy (TEM) and SEM (Fig. 1e-f), as well as by atomic force microscopy (AFM) with which we have estimated the nominal thickness of the exfoliated 2D sheets is of the order of 10s nm (Fig. 1g). Additional microscopy images showing the 2D nanosheet morphologies are presented in the Supporting Information (SI), see figs. S1-S12.

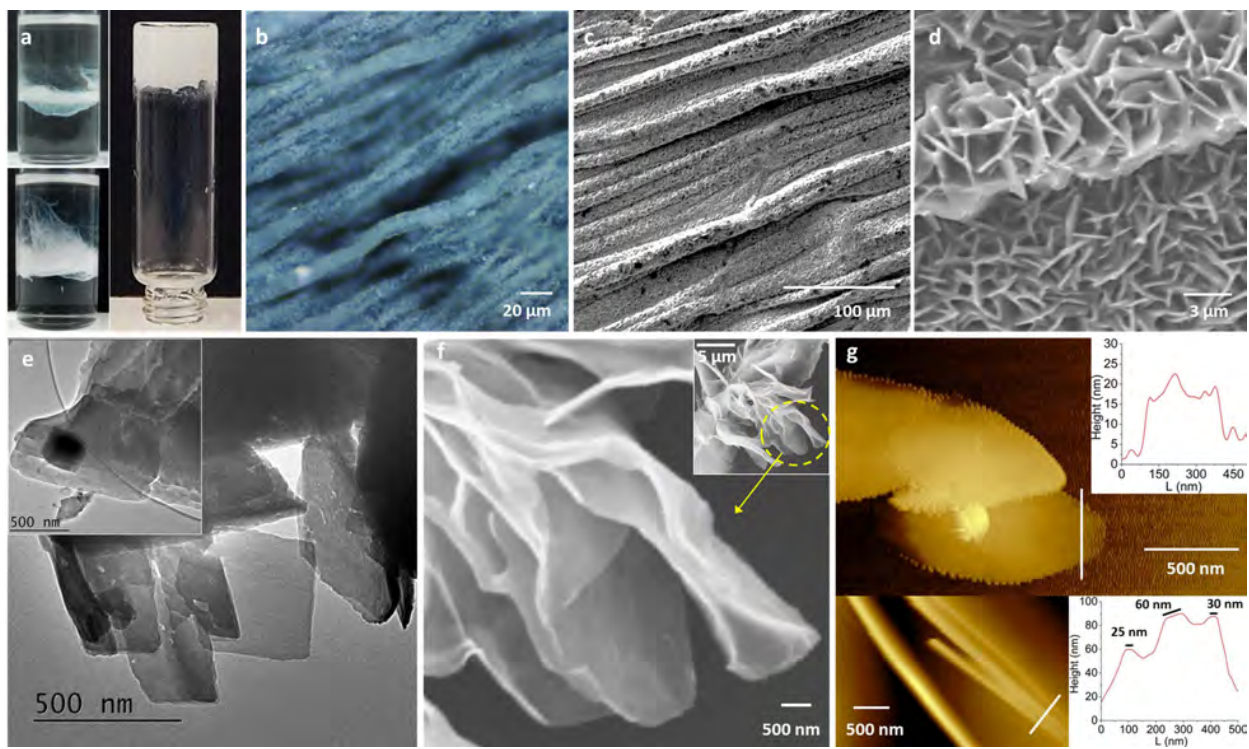

**Fig. 1. Two-dimensional nanosheets of self-assembled OX-1 MOF materials.** (a) Supramolecular synthesis showing stepwise development of fibrous soft matter (left), forming a bulk of hybrid gel material (right: confirmed by vial inversion test). (b) Optical microscopy image of highly oriented fibrous material revealing shiny fiber surfaces. (c) SEM images of aligned fiber microstructures, and (d) densely packed nanosheets on gel fibers connected by weak intermolecular interactions. Samples (b-d) were synthesized by layering reactant solutions onto a flat glass substrate. (e) TEM images of thin 2D morphologies of the exfoliated OX-1 nanosheets; the inset shows similar morphology of functionalized (Guest@Host) nanosheets of $ZnQ_{DMF}@OX-1$. (f) SEM of $ZnQ_{DMF}@OX-1$ nanosheets revealing the dissociated 2D layers (this image is a magnified view of the inset micrograph) (g) AFM height topography showing the thickness of exfoliated layers.

Powder X-ray diffraction studies (PXRD) (Fig. 2a-b) of nanosheets extracted from the supramolecular fibers (Fig. 1e-f) showed crystallographic resemblance to a 3D MOF structure: $(H_2NEt_2)_2[Zn_3BDC_4]\cdot3DEF$ reported by Burrows *et. al*[15] and Stock *et. al*[16]. However, there are important differences between the present MOF nanosheets and the materials mentioned earlier, from both the structural and the synthetic points of view. First, our diffraction data evidenced strong signature of peak shifts, especially of the diagonal (110) planes (Fig. 2b), which we solved by Pawley refinement (Fig. 2a) revealing salient variations in the basic unit cell geometry (triclinic *vs*. monoclinic, see table S1); and there are substantial peak broadening arising from the fine-scale nanosheet morphology. Crucially the crystal structure has predominantly unidirectional pores (Fig. 2c), comprising 1D undulating channels down the *c*-axis. Second, neutralization of the framework negative charges was previously mediated by diethylammonium cations ($H_2NEt_2^+$) liberated from hydrolysis of *N*,*N*-diethylformamide (DEF) solvent, accomplished either in a strong acidic solution (2M $HNO_3$) at high-temperature synthesis (120°C, 1 day)[16], or, in the presence of water molecules under protracted conditions (~weeks)[15]. In contrast, MOF nanosheets prepared *via* our rapid HCR method at room temperature (less than 1 min) incorporate charge-balancing cations $NEt_3^+$ (derived from its neutral form, when activating BDC linkers). Considered together our results show that, inclusion of $NEt_3^+$ cations inside MOF pores leads to deformation of host framework, where structural distortion by straining has reduced unit cell symmetry (table S1). We subsequently established the chemical formula of the present MOF structure to be: $(HNEt_3)_2[Zn_3BDC_4] \cdot solvent$ where *solvent* = DMF or DMA (*vide infra*), validated by thermogravimetric analysis (TGA) (fig. S16-S17). Hereafter the identified MOF nanosheet structure is designated as "OX-1" (*i.e.* Oxford University-1 material).

Remarkably our HCR strategy can be adopted, to directly functionalize OX-1 nanosheets to derive new photoactive "Guest@Host" composite systems. To illustrate this, we demonstrate *in situ* nanoscale confinement of the luminescent metal complex "guest" molecule: zinc(II) bis(8-hydroxyquinoline)[17], termed ZnQ, spatially confined inside the 1D pore channels of the OX-1

"host" framework employing one-pot supramolecular synthesis[14]. We discovered that, the coupled Guest@Host systems synthesized from $N,N$-dimethylformamide (DMF) and $N,N$-dimethylacetamide (DMA) solvents displayed significantly different luminescent behavior, although identical synthetic conditions[14] were applied, other than their solvent type. When subject to UV irradiation it can be seen that (Fig. 2d), nanosheets synthesized in DMF solvent emit an intense blue light ($\lambda_{em}$ = 470 nm), whereas samples from DMA solvent display a green light emission ($\lambda_{em}$ = 510 nm). Henceforth, we designate these two new Guest@Host nanosheet systems as: $ZnQ_{DMF}$@OX-1 and $ZnQ_{DMA}$@OX-1, respectively (subscript denotes solvent type used in synthesis). To understand their differential structure-property relationships, we performed spectroscopic measurements to study the detailed photophysical properties.

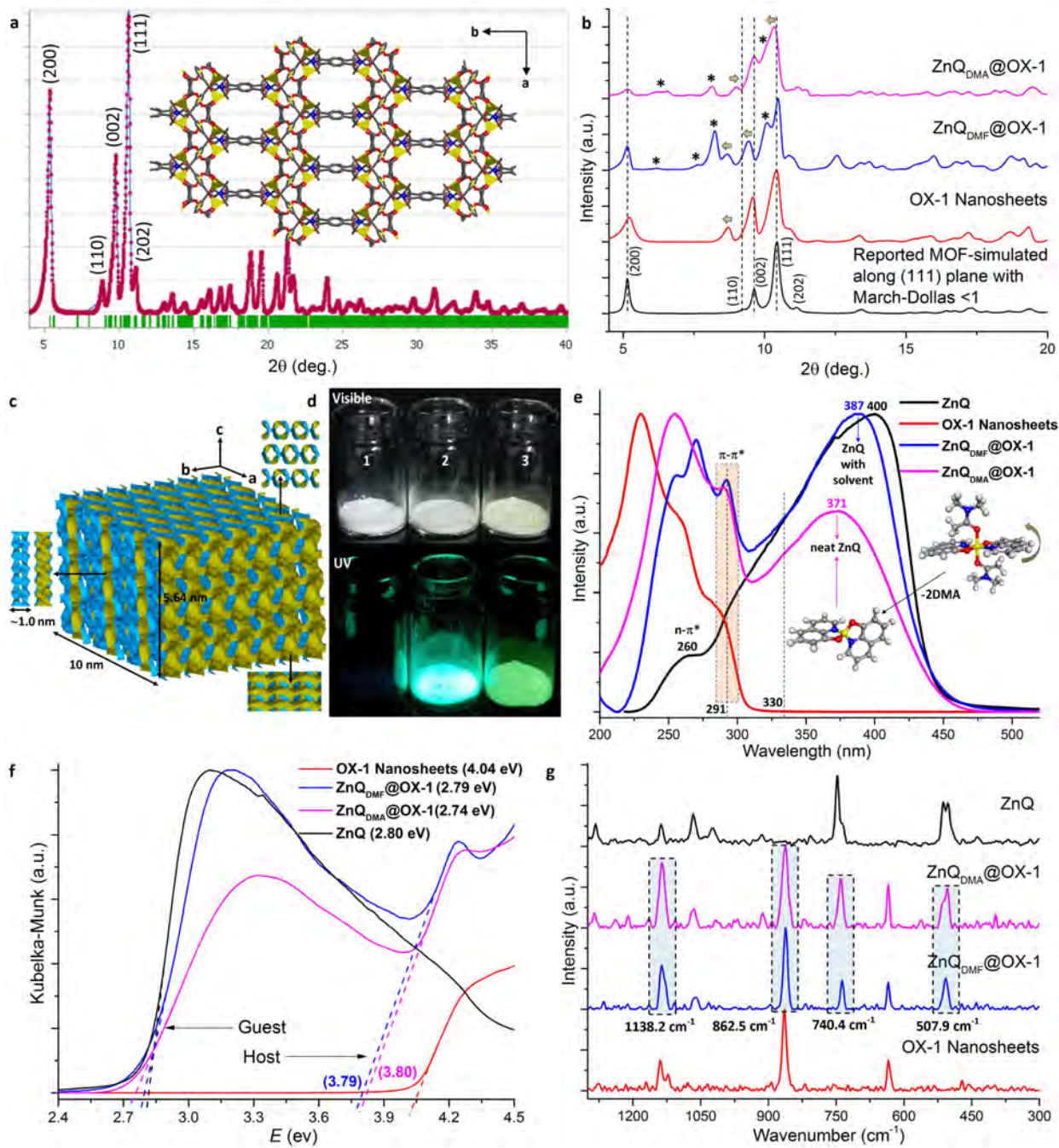

**Fig. 2. Structural and spectroscopic characterization of the OX-1 nanosheets and their functionalized Guest@OX-1 counterparts.** (a) Pawley refinement[18] of the OX-1 crystal structure incorporating $NEt_3^+$ charge-balancing cations ($R_{wp} = 0.084$), and crystallographic view along the *c*-axis showing 1D porosity (inset). Color scheme: zinc in yellow, nitrogen in blue, oxygen in red, carbon in gray, hydrogen in white. The traces indicate data as follows; red: experimental data, blue: calculated from structural refinement, green: observed Bragg peak positions (b) PXRD of nanosheets, with and without confinement of the ZnQ guest molecules, compared with the simulated diffraction pattern from ref.[15]. Asterisks mark the positions of the

ZnQ peaks. (c) Undulating pore architecture of 1D channels along *c*-axis, solvent accessible volume is denoted by yellow surfaces. (d) Emission tests under 365 nm UV irradiation, for sample 1: pristine OX-1 nanosheets without ZnQ is non-emissive; 2: $ZnQ_{DMF}$@OX-1 and 3: $ZnQ_{DMA}$@OX-1 are optically active. (e) UV-Vis electronic absorption spectra of the OX-1 host framework, ZnQ guest emitter, and guest@OX-1 composite systems. Inset depicts conformational changes of the neat ZnQ guest emitter molecule, after losing two DMA coordinated solvent molecules from its axial positions. (f) Band gap values determined from diffuse reflectance spectra and photon energy intercepts, signifying host-guest interactions (g) Raman spectra revealing intensity and peak alterations arising from the host-guest confinement effects.

**Photophysical and photochemical properties of Guest@MOF nanosheets**

Absorption UV-Vis spectroscopic measurements were used to elucidate the nature of the host-guest interactions and to identify spatial confinement effects due to nanoscale porosity. It can be seen in Fig. 2e that, functionalized nanosheets of $ZnQ_{DMF}$@OX-1 and $ZnQ_{DMA}$@OX-1 exhibit an appreciably enhanced π—π* electronic transition at 291 nm, confirming the successful encapsulation of luminous ZnQ guest emitter within the pores of the OX-1 host framework. This energy transfer phenomenon is caused by close-packing of host-guest aromatic moieties, characteristic of the caged guest molecules. Fig. 2f shows the modification of band gaps as a consequence of intimate host-guest coupling. Through encapsulation of ZnQ in OX-1 pores we have evidenced a clear reduction in band gap of the OX-1 framework, which fell from greater than 4 eV to just below 2.8 eV. Interestingly, we found that $ZnQ_{DMA}$@OX-1 exhibits a maximum absorption corresponding to the ZnQ guest contribution at 371 nm, indicative of the confinement of *neat* ZnQ[19] inside OX-1 pores. Compared to DMF, because DMA is a relatively bulkier molecule, its coordination to the Zn center of ZnQ at two axial positions (Fig. 2e inset) might be misplaced upon pore confinement, implicated by caging in the spatially constrained channels (Fig.3). Neat ZnQ complex thus confers extra degrees of freedom to the coordinated bis-8-hydroxyquinoline (8HQ) aromatic moieties around the $Zn^{2+}$ center of the guest emitter; this effect is illustrated in the inset of Fig. 2e. In contrast, because Zn metal centers strongly prefer DMF over DMA[20], coordinated DMF helps to stabilize the ZnQ structure. This is in agreement with the absorption data of $ZnQ_{DMF}$@OX-1 nanosheets, where $\lambda_{max}$ was pinpointed at 387 nm indicating that the ZnQ has coordinated DMF.

Utilizing Raman vibrational spectroscopy (Fig. 2g), we achieved further insights into symmetry alterations of the ZnQ guest emitter as affected by pore confinement. For $ZnQ_{DMA}$@OX-1, the doubly-degenerate Raman modes at 503.8 cm$^{-1}$ (assigned to skeletal in-plane bending vibrations[21]) and 514.04 cm$^{-1}$ (fig. S18) are pointing towards a reduction in molecular symmetry of the neat ZnQ (*i.e.* without DMA coordination, see Fig. 2e inset); this phenomenon is prominent also in pure ZnQ. Interestingly for $ZnQ_{DMF}$@OX-1, only a single band was identified at 507.9 cm$^{-1}$, which meant that the confined ZnQ guest (with DMF) has higher structural symmetry. Relative mode intensities at 740.42 cm$^{-1}$ associated with aromatic ring deformation is appreciably higher in $ZnQ_{DMA}$@OX-1 compared to $ZnQ_{DMF}$@OX-1, further supporting the notion that the former nanosheet confines neat ZnQ (higher degrees of freedom). The band at 862.5 cm$^{-1}$ corresponds to the out-of-plane $\gamma$(C-H) bending mode of BDC molecules[22] in the host framework. For $ZnQ_{DMF}$@OX-1, this mode is accompanied by a doubly-degenerate band identified at a lower frequency of 852.2 cm$^{-1}$ (fig. S18), implicated by strong intermolecular interactions of host-guest aromatic rings hence lessening structural symmetry.

We measured the fluorescence quantum yield (QY%) of pure ZnQ guest emitter in DMF suspension (35%) and found that, it has risen to 43.8% upon nanoconfinement in the $ZnQ_{DMF}$@OX-1 system, which further resulting in an improved emission lifetime (fig. S19). Conversely, QY of the $ZnQ_{DMA}$@OX-1 system deteriorated to 23.3% (compared to 33% for pure ZnQ in a DMA suspension, see fig. S19). Our findings are in line with the reported photophysical phenomena where ground and excited states of metal complexes can be modified by fine structural changes from pressure effects, which are manifested as decreasing non-radiative or increasing radiative rates[23]. Our data suggest that ZnQ caged inside ZnQ$_{DMF}$@OX-1 is experiencing compressive strains enhancing its luminescence emission; but ZnQ$_{DMA}$@OX-1 is showing a reduction in luminescence due to higher intermolecular interactions between neat ZnQ aromatic rings and the host framework. This reasoning is consistent with the UV-Vis and Raman spectroscopic data discussed above.

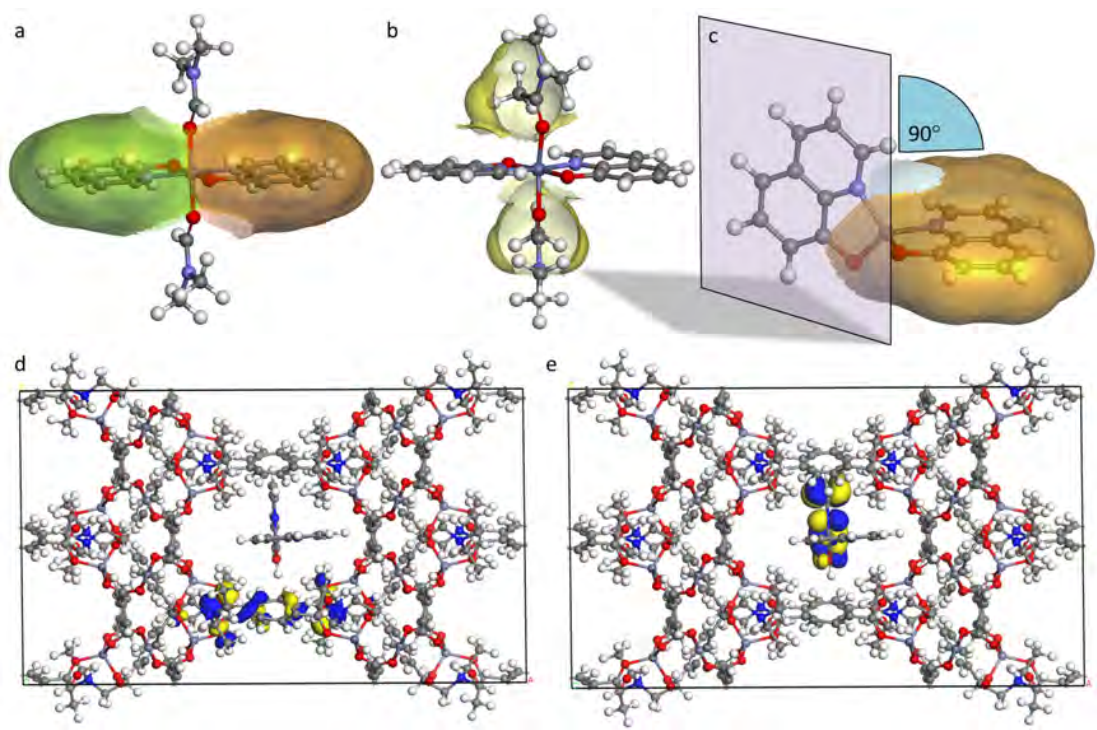

**Fig. 3. Theoretical determination of the guest molecular structure and electronic orbitals associated with host-guest interactions.** Optimized geometries of the ZnQ guest in three different structural configurations: (a) DMF solvent molecules coordinated axially to the Zn$^{2+}$ octahedral center; less steric DMF molecule allowing planar spatial arrangement of the 8HQ aromatic rings (indicated by green and orange van der Waals surfaces). (b) Steric hindrance from axially coordinated DMA solvent molecules deforming the planarity of 8HQ aromatic rings; yellow surfaces represent CH$_3$ groups of the coordinated DMA. (c) Tetrahedral (Td) configuration achieved upon removal of the two axially coordinated DMA molecules, permitting a 90° rotation of one of the two 8HQ aromatic planes. ZnQ(Td)@OX-1 assembly in which (d) HOMO located on the BDC linker of the host framework, while (e) LUMO on the aromatic rings

of ZnQ guest molecule confined within OX-1 pore. Blue and yellow isosurfaces are positive and negative charges respectively; see further details in SI.

Our experimental results are further supported by theoretical calculations (see §10 in SI), elucidating that the ZnQ guest molecule could adopt different geometrical configurations under pore confinement. Fig.3a shows the optimized structure of ZnQ with two DMF molecules coordinated to the $Zn^{2+}$ axial positions, where the adjacent 8HQ aromatic rings are planar. However, these rings became non-planar when DMF was substituted by the bulkier DMA molecules (Fig.3b), leading to a significant geometrical distortion in confinement of the OX-1 framework (fig. S21); such a host-guest configuration is not favorable. In fact, our calculations revealed that ZnQ guest with tetrahedral center (Td) has the preferred geometry (Fig.3c: without coordinated DMA) formed by a 90° rotation of the 8HQ aromatic plane; this configuration also offers good interactions with the host framework *via* π-π and hydrogen bonding (fig. S23).

Density functional theory (DFT) band gap calculations show that, for example, when considering ZnQ(Td)@OX-1 pore confinement (Fig. 3d,e) the highest occupied molecular orbital (HOMO) is localized on the BDC ring of the OX-1 host, whereas the lowest unoccupied molecular orbital (LUMO) is located on the aromatic rings of the ZnQ molecule. This HOMO-LUMO configuration suggests that host-guest energy transfer will be favorable during photoexcitation of the Guest@Host assembly. Detailed modifications of the electronic energy levels as a consequence of ZnQ guest confinement in the pore of the OX-1 host are summarized in fig. S24, and their HOMO-LUMO frontier orbitals are presented in table S2. It can be seen that the hybrid orbitals of the ZnQ(Td)@OX-1 and $ZnQ_{DMF}$@OX-1 systems are exhibiting a notable reduction in bandgap energies, relative to the isolated OX-1 host and pure ZnQ guest molecule. Our results further support the recent literature where it has been proposed that: (i) mechanical deformation of flexible MOF structures could produce significant band gap changes[24] and, (ii) localized electronic density distribution of an encapsulated host-guest assembly could enable efficient energy transfer during excitation process.[25]

**Optochemical sensing applications**

We harnessed the tunable luminescent properties of the functionalized Guest@OX-1 nanosheets as photactive materials, accomplishing optochemical sensing of volatile organic compounds (VOCs). Fig. 4a and fig. S26 show the luminescence data of $ZnQ_{DMF}$@OX-1 nanosheets as solution-state dispersions, upon perturbation by a series of main solvent species used as analytes. We discover major photophysical response not only in characteristic fluorescence wavelength and intensity (Fig. 4b), but also in terms of color chromaticity tuning behavior (Fig. 4c). Between the small aliphatic alcohols we probed, methanol caused a fully diminished fluorescence intensity (quenching) after 3-4 minutes exposure, while nanosheets in isopropanol (IPA) retained majority of its intensity. In fact with rising alcohol polarity (fig. S27), we observed a sharper decline in emission intensity and a higher frequency red shift following the sequence of: methanol (510 nm) > ethanol (504 nm) > IPA (501 nm). Strongly polarizable protic solvent destabilizes the ZnQ emitter guests, leading to fluorescence quenching *via* protonation of coordinated oxygen in 8HQ linkers[26]. Our data indicate excited state proton-induced charge transfer of ZnQ guests in protic solvents[27], where such a mechanism depresses luminescence. The optochemical sensing results collected from a systematic study of 14 example solvent molecules and VOCs are summarized in fig. S26-S27.

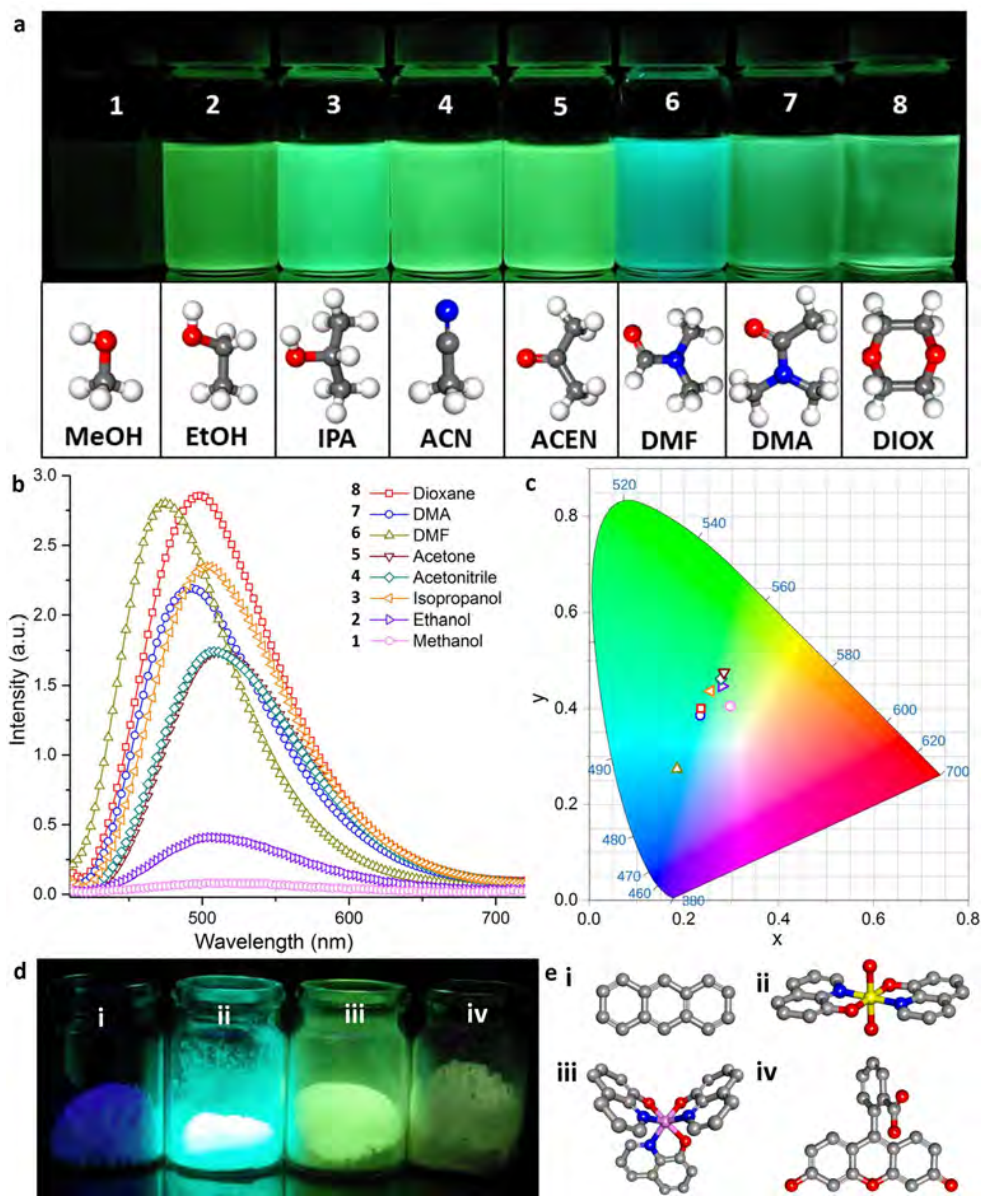

**Fig. 4**. **Tunable optochemical behavior of ZnQ$_{DMF}$@OX-1 nanosheets.** (a) Visible to naked eye are distinct modulations in emission properties of dispersions of functionalized nanosheets in a range of small-molecule solvents (each comprising 5 mg of active material in 15 mL solvent). (b) Emission profiles and emission intensities of respective dispersions showing different levels of blue (hypsochromic) or red (bathochromic) shifts. (c) Chromaticity plot (CIE 1931) indicating the emission color coordinates of the respective dispersions. (d) A family of luminescent host-guest materials synthesized by one-pot supramolecular method (Scheme 1), adopting the same OX-1 host framework but confining different light-emitting guest molecules (e): 1 - Anthracene, 2 - ZnQ, 3 - AlQ (Al-(tris-8-Hydroxyquinoline), 4 - Fluoresceine. Color scheme: zinc in yellow, nitrogen in blue, oxygen in red, carbon in gray, hydrogen in white.

Functionalized nanosheets exposed to non-polar aliphatic long-chain alkane (*n*-hexane) and cyclic alkane (cyclohexane) displayed interesting optochemical stimulation behaviors. Samples of ZnQ$_{DMF}$@OX-1 ($\lambda_{em}$ = 470 nm) dispersed in these aliphatic analytes produced similar fluorescence intensities, but peaked at very different characteristic wavelengths ($\lambda_{em(n\text{-hexane})}$ = 496 nm *vs.* $\lambda_{em(cyclohexane)}$ = 477 nm, see fig. S26). This meant that higher molecular mobility of linear *n*-hexane in the OX-1 pores generates a stronger red shift, which can be used to distinguish against the bulkier cyclohexane analyte. Indeed molecular-size selectivity effect was best demonstrated for the aromatic pair — benzene *vs.* toluene (both of similar polarity), where nanosheets exposed to bulkier toluene experienced only a minor intensity rise, however that caused by benzene was the most intense detected amongst all the analytes we tested (fig. S26). Significantly this result elucidates that by increasing π—π interaction within the ZnQ$_{DMF}$@OX-1 system, it strengthens fluorescence intensity without modifying the emission wavelength. Considering yet another pair of related analytes — dichloromethane (DCM) *vs.* chloroform, we established that the relatively smaller DCM molecules can readily infiltrate the tortuous 1D MOF channels (Fig. 2c) yielding a major red shift (>30 nm), while diminishing fluorescence intensity due to its strong electron withdrawing nature. Collectively our comprehensive perturbation data confirmed that, the functionalized nanosheets are extremely sensitive towards salient solvent parameters (fig. S27), encompassing polarity, molecular size, hydrogen bond donor and acceptor affinity, and non-covalent π—π interactions. Long-range crystallinity of the OX-1 host implies that there is a well-ordered arrangement of functional guests (e.g. ZnQ emitters), where its periodicity and accessible porosity (fig. S25) offers direct interaction with incoming solvent analytes permitting fast chemical detection. In fact we must recognize that, analyte dependent sensing capability is absent when deploying pure ZnQ complexes alone (fig. S28), *i.e.* without the host-guest nanoconfinement environment imparted by the porous OX-1 nanosheets.

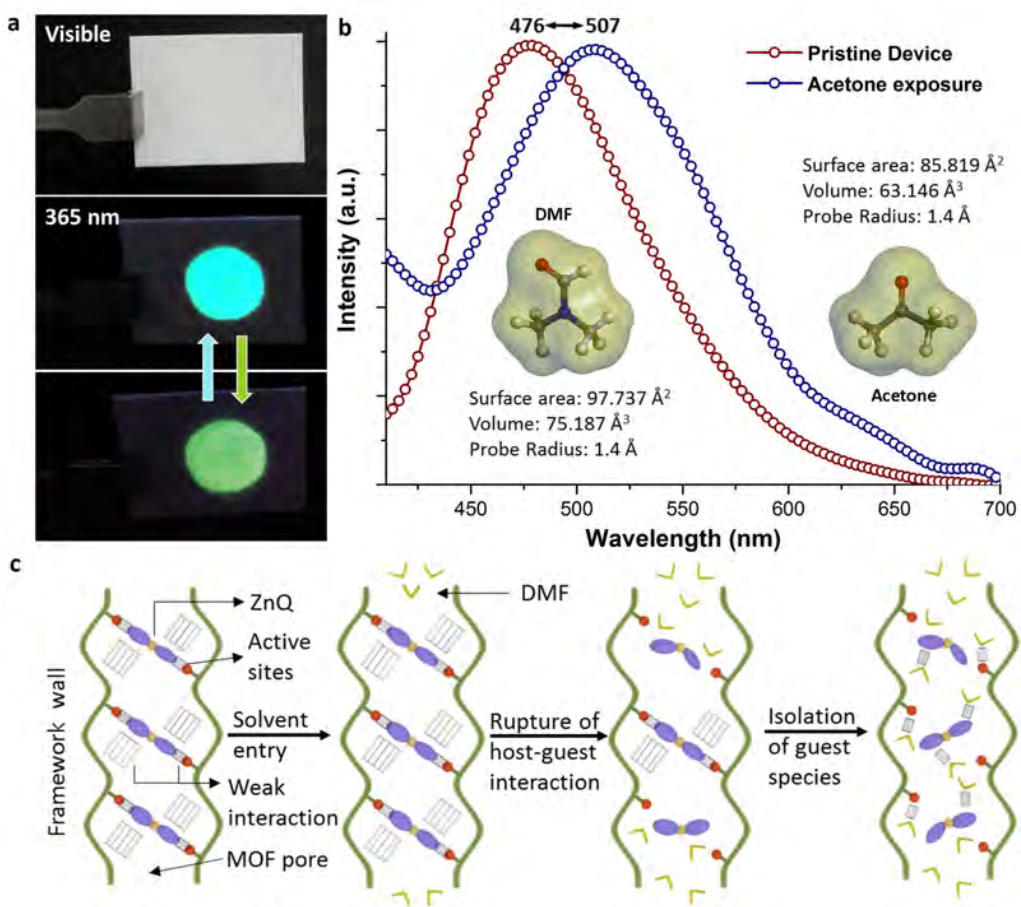

**Fig. 5. Optochemically active sensory materials comprising ZnQ$_{DMF}$@OX-1 nanosheets.** (a) Paper-based optochemical sensor demonstrating reversible acetone sensing ability. (b) A 31-nm red shift upon acetone exposure and, recovery to its original emission wavelength upon DMF exposure. Inset shows chemical structures of DMF and acetone analytes and their differential molecular sizes. (c) Proposed mechanisms involving multiple host-guest interactions under the nanoscale confinement of 1D undulating (tortuous) channels of OX-1 host framework (vertical: *c*-axis, corresponding to 1D channels in Fig.2c), and upon the introduction of a DMF analyte triggering a blue shift transformation.

Motivated by the capacity exhibited by functionalized guest@OX-1 nanosheets in the detection of numerous solvent analytes and VOCs, we present a simple proof-of-concept device demonstrating the efficacy of ZnQ$_{DMF}$@OX-1 as a reversible solid-state chemical sensor. A paper-based device (Fig. 5a) was fabricated by drop coating a small amount of ZnQ$_{DMF}$@OX-1 (*e.g.* ~2 µL dispersion made from 5 mg nanosheets suspended in 1 mL hexane) onto a Whatman filter paper. We tested this paper device with a small concentration of analyte. For example, ~1 µL acetone applied to this paper device swiftly produced a substantial 31 nm red shift (under 365 nm UV), which is clearly perceptible to the naked eye (Fig. 5a). Intriguingly this transition is reversible (507 ⇌ 476 nm, Fig. 5b); the device is again reusable after ~10 sec drying time at room temperature as acetone is volatile (in contrast we note that ZnQ$_{DMA}$@OX-1 has an initial emission wavelength of $\lambda_{em}$ = 510 nm, thus rendering it unsuitable for wavelength-shift based sensing of acetone). Nonetheless, we found that repetitive exposures to acetone could retain the device emission at a higher wavelength of 507 nm. This finding suggests that increased acetone occupancy in OX-1 pores is strengthening non-covalent interactions bridging the ZnQ emitter to the host framework. However, even such an effect is non-permanent in nature; it can be reverted to its initial 476 nm (blue) emission simply by exposing the device to DMF for several minutes. Key observations made in emission response upon DMF exposure let us elucidate the plausible host-guest interaction mechanisms, as depicted in Fig. 5c. The moment the active material was exposed to DMF, it was observed that emission color quickly turned greenish-yellow for fraction of a second before beginning to undergo blue shifting and eventually returning to its original 476 nm. On this basis, we hypothesize that weak interactions connecting the ZnQ guest and the host framework sites began to slowly rupture upon infiltration of external DMF analyte. We reasoned that penetration of DMF molecules through the 1D tortuous pore channels of the OX-1 host is relatively sluggish, thereby isolating ZnQ from adjacent weak interactions in a stepwise fashion (Fig. 5c); eventually this reaches a complete conversion evidenced by the (blue) 476 nm emission. Comparatively immediate response towards acetone exposure can be explained by contrasting the molecular size of the analytes concerned (Fig. 5b inset): acetone molecule is approximately 16% smaller compared to DMF, so the former infiltrates the undulating pores much easier given its smaller size and weakly interacting nature.

In summary, we have elucidated a new paradigm to prepare Guest@MOF porous nanosheets, resulting in tunable nanoscale material systems highly sensitive to coupled optoelectronic and chemical perturbations. The proposed high-concentration reaction (HCR) supramolecular strategy is very powerful, and it could be utilized to systematically engineer a variety of functionalized Guest@MOF composite systems. To show that this is an exciting possibility, we have implemented the HCR approach in conjunction with the Guest@OX-1 platform to integrate a number of topical guest emitters successfully (see Figs. 4d-e), including anthracene, ZnQ, AlQ [Al-(tris-8-Hydroxyquinoline], naphthalene, and fluorescein[14]. In the light of this, we anticipate that the proposed methodology will open the door to a new family of low-dimensional MOF-based smart materials, whose bespoke physico-chemical properties will be of significant utility to the chemical sensors, photonics, and electronics material sectors.

**Methods**

**Synthesis of porous OX-1 2D nanosheets as host framework material:** 2 mL *N,N*-dimethylformamide (DMF) clear solution of $Zn(NO_3)_2$ (1.5 mmol, termed Solution A) was combined with 3 mL clear solution of 1,4-benzenedicarboxylic acid (3 mmol, BDC) plus triethylamine ($NEt_3^+$) (6 mmol) (termed Solution B). A fibrous gel-like supramolecular material formed immediately upon combining solutions A and B. The gel phase is disordered, comprising both crystalline (2D nanosheets) and non-crystalline materials (Scheme 1c) in addition to excess amounts of reactants and guest molecules. The nanosheet-rich fibers were washed twice with copious amounts of polar solvents: first using DMF followed by methanol and acetone, and with simultaneous sonication (10 minutes per solvent, then centrifugation). This washing procedure quickly breaks down the supramolecular fibers to release the 2D nanosheets (insoluble), and at the same time removing externally adhered guest species (not in MOF pores) and excess reactants. Finally, nanosheets harvested can be separated by centrifugation (8000 rpm) and subject to vacuum drying at 110 °C for 4 hours. The isolated nanosheets are highly crystalline MOF material: $(HNEt_3)_2[Zn_3BDC_4]\cdot DMF$, designated as "OX-1" nanosheets.

**Synthesis of functionalized MOF nanosheets of $ZnQ_{DMF}@OX-1$ and $ZnQ_{DMA}@OX-1$:** Preparation of functionalized MOF nanosheet materials (Guest@OX-1) involves a simple additional step of *in situ* mixing of *N,N*-Dimethylformamide (as a clear solution) containing the desired guest species (*e.g.* ZnQ/AlQ/Naphthalene/Anthracene/Fluorescein) into Solution B, see §1.2 in SM for details. The procedures for isolating, washing, and harvesting nanosheets are identical to those described in the foregoing section.

**Materials characterization:** The fiber architecture of supramolecular MOF material was confirmed using optical microscopy (Alicona Infinite Focus 3D microscope) and scanning electron microscopy (SEM, Carl Zeiss EVO LS15). Detailed nanosheet structures and morphologies were examined under SEM and transmission electron microscopes (TEM, JEOL JEM-2100 LaB6, 200 kV), and atomic force microscopy (AFM, Veeco Dimension 3100). Kubelka-Munk (K-M) function was derived from diffuse reflectance spectra measured by UV-2600 UV-Vis spectrophotometer, Shimadzu. Steady-state emission spectra and CIE 1931 were recorded using UPRtek spetrophotometer (MK350N Plus). Raman spectroscopy was performed using MultiRAM FT-Raman Spectrometer (Bruker) equipped with a 532 nm laser. Powder X-ray diffraction (PXRD) pattern was recorded using the Rigaku MiniFlex with a Cu Kα source (1.541 Å), where diffraction data were collected at 2$\theta$ angle from 2° to 30°, using a 0.01° step size and 1° min$^{-1}$ step speed. Quantum yield and fluorescence lifetime measurements were performed on Quantaurus-QY Absolute PL quantum yield spectrometer (C11347) and Quantaurus-Tau fluorescence lifetime spectrometer (C11367), respectively. Diluted nanosheet dispersions in respective solvents were used for fluorescence QY and lifetime measurements.

---

**Acknowledgments:** This research is funded by the Samsung Advanced Institute of Technology (SAIT) GRO, the Royal Society Research Grant (RG140296), and the Engineering and Physical Sciences Research Council, EPSRC RCUK (EP/N014960/1 and EP/K031503/1). We thank the Research Complex at Harwell (RCaH), Oxfordshire, for access to advanced materials characterization suite. We are grateful to Dr. Gavin Stenning and Dr. Marek Jura (R53 Materials Characterization Lab) at the ISIS Rutherford Appleton Laboratory, for provision of X-ray diffraction facilities. We acknowledge Dr. James Taylor (R79 Hydrogen and Catalysis Laboratory at ISIS) for performing the BET measurements. We thank Dr. M.J. Abdin from Hamamatsu Photonics Ltd. for the provision of quantum yield and fluorescence lifetime spectrometers.

**Author Contributions:** AKC and JCT designed the research. AKC performed the experiments and analyzed the data guided by JCT. HJK and ITH contributed to quantum yield measurements and technical insights into sensor device applications. AKC and JCT wrote the paper, with input from all authors.

**References**

[1]  a) H. L. Peng, et al., *Nat. Chem.* **2012**, *4*, 281-286; b) M. Osada, T. Sasaki, *Adv. Mater.* **2012**, *24*, 210-228.

[2]  a) J. Choi, H. Y. Zhang, J. H. Choi, *ACS Nano* **2016**, *10*, 1671-1680; b) J. N. Coleman, et al., *Science* **2011**, *331*, 568-571; c) J. C. Tan, P. J. Saines, E. G. Bithell, A. K. Cheetham, *ACS Nano* **2012**, *6*, 615-621.

[3]     a) Y. Peng, et al., *Science* **2014**, *346*, 1356-1359; b) K. Varoon, et al., *Science* **2011**, *334*, 72-75.

[4]     A. C. Ferrari, et al., *Nanoscale* **2015**, *7*, 4598-4810.

[5]     a) J. M. Lehn, *Proc. Natl. Acad. Sci. USA* **2002**, *99*, 4763-4768; b) E. R. Draper, E. G. B. Eden, T. O. McDonald, D. J. Adams, *Nat. Chem.* **2015**, *7*, 849-853.

[6]     B. Choi, et al., *Nano Lett.* **2016**, *16*, 1445-1449.

[7]     a) G. Xu, et al., *J. Am. Chem. Soc.* **2012**, *134*, 16524-16527; b) Y. Sakata, et al., *Science* **2013**, *339*, 193-196; c) M. Zhao, et al., *Adv. Mater.* **2015**, *27*, 7372-7378; d) T. Rodenas, et al., *Nat. Mater.* **2015**, *14*, 48-55; e) C. Hermosa, et al., *Chem. Sci.* **2015**, *6*, 2553-2558.

[8]     a) A. J. Howarth, et al., *Nat. Rev. Mater.* **2016**, *1*, 15018; b) B. Garai, A. Mallick, R. Banerjee, *Chem. Sci.* **2016**, *7*, 2195-2200; c) M. S. Yao, et al., *Adv. Mater.* **2016**, *28*, 5229-5234; d) M. R. Ryder, J. C. Tan, *Mater. Sci. Tech.* **2014**, *30*, 1598-1612.

[9]     a) R. Ameloot, et al., *Nat. Chem.* **2011**, *3*, 382-387; b) M. O'Keeffe, O. M. Yaghi, *Chem. Rev.* **2012**, *112*, 675-702; c) M. R. Ryder, B. Civalleri, G. Cinque, J. C. Tan, *CrystEngComm* **2016**, *18*, 4303-4312.

[10]    a) N. Liedana, et al., *ACS Appl. Mater. Interfaces* **2012**, *4*, 5016-5021; b) R. Anand, et al., *J. Phys. Chem. B* **2014**, *118*, 8532-8539; c) A. K. Chaudhari, M. R. Ryder, J. C. Tan, *Nanoscale* **2016**, *8*, 6851-6859; d) W. Xie, et al., *Chem. Commun.* **2016**, *52*, 3288-3291.

[11]    a) M. D. Allendorf, et al., *J. Phys. Chem. Lett.* **2015**, *6*, 1182-1195; b) J. Aguilera-Sigalat, D. Bradshaw, *Coord. Chem. Rev.* **2016**, *307*, 267-291.

[12]    a) L. E. Kreno, et al., *Chem. Rev.* **2012**, *112*, 1105-1125; b) Z. Hu, B. J. Deibert, J. Li, *Chem. Soc. Rev.* **2014**, *43*, 5815-5840.

[13]    A. K. Chaudhari, I. Han, J. C. Tan, *Adv. Mater.* **2015**, *27*, 4438-4446.

[14]    See supplementary materials - Materials and Methods

[15]    A. D. Burrows, et al., *CrystEngComm* **2005**, *7*, 548-550.

[16]    E. Biemmi, T. Bein, N. Stock, *Solid State Sci.* **2006**, *8*, 363-370.

[17]    L. S. Sapochak, et al., *J. Am. Chem. Soc.* **2002**, *124*, 6119-6125.

[18]    G. S. Pawley, *J. Appl. Crystallogr.* **1981**, *14*, 357-361.

[19]    N. Khaorapapong, M. Ogawa, *J. Phys. Chem. Solids* **2008**, *69*, 941-948.

[20]    S. Ishiguro, Y. Umebayashi, R. Kanzaki, *Anal. Sci.* **2004**, *20*, 415-421.

[21]    V. Krishnakumar, R. Ramasamy, *Spectrochim. Acta A* **2005**, *61*, 673-683.

[22]    M. W. Lee, M. S. Kim, K. Kim, *J. Mol. Struc.* **1997**, *415*, 93-100.

[23]    a) R. W. Parsons, H. G. Drickamer, *J. Chem. Phys.* **1958**, *29*, 930; b) J. K. Grey, I. S. Butler, *Coord. Chem. Rev.* **2001**, *219*, 713-759; c) C. Reber, *Can. J. Anal. Sci.* **2008**, *53*, 91-101.

[24]    S. Ling, B. Slater, *J. Phys. Chem. C* **2015**, *119*, 16667-16677.

[25]    D. Yan, Y. Tang, H. Lin, D. Wang, *Sci. Rep.* **2014**, *4*, 4337.

[26]    B. Valeur, I. Leray, *Coord. Chem. Rev.* **2000**, *205*, 3-40.

[27]    E. Bardez, I. Devol, B. Larrey, B. Valeur, *J. Phys. Chem. B* **1997**, *101*, 7786-7793.

**Supporting Information**

**for**

**Optochemically Responsive 2D Nanosheets of a
3D Metal-Organic Framework Material**

*Abhijeet K. Chaudhari,*[1] *Ha Jin Kim,*[2] *Intaek Han,*[2] and *Jin-Chong Tan*[1*]

[1]*Department of Engineering Science, University of Oxford, OX1 3PJ, Oxford,
United Kingdom*
[2]*Samsung Advanced Institute of Technology (SAIT), Samsung Electronics Co. Ltd.,
South Korea*

*Correspondence to: jin-chong.tan@eng.ox.ac.uk

**Table of Contents**

# 1    Materials and Methods

## 1.1    Synthesis of Porous OX-1 2D Nanosheets as Host Framework Material

2 mL $N,N$-Dimethylformamide (DMF) clear solution of $Zn(NO_3)_2$ (1.5 mmol, termed Solution A) was combined with 3 mL clear solution of 1,4-Benzenedicarboxylic acid (3 mmol, BDC) plus Triethylamine ($NEt_3^+$) (6 mmol) (termed Solution B). We found fibrous hybrid materials formed immediately upon combining solutions A and B.

The hybrid fiber material containing nanosheets was washed with copious amounts of DMF, methanol and acetone to break down the gel fibres, thereby releasing pure nanosheets of porous MOF material: $(HNEt_3)_2[Zn_3BDC_4]\cdot DMF$ (we designate as "OX-1" MOF nanosheets). Further sonication step can be used to exfoliate the OX-1 material into even thinner 2D sheets, which suggests the presence of weak molecular interactions between the adjacent 3D MOF structure hence permitting facile 2D exfoliation. Finally, the nanosheets harvested were separated by centrifugation (8000 rpm) and then subjected to vacuum drying at 110 °C for 4 hours.

## 1.2 Synthesis of Functionalized MOF Nanosheets: ZnQ_DMF@OX-1 and ZnQ_DMA@OX-1

Preparation of functionalized MOF nanosheet materials (Guest@OX-1) involves a simple additional step of *in situ* mixing of *N,N*-Dimethylformamide (as clear solution) containing guest species *e.g.* ZnQ/AlQ/Naphthalene/Anthracene/Fluorescein into Solution B (see §1.1). This facile high-concentration reaction (HCR) method is shown below.

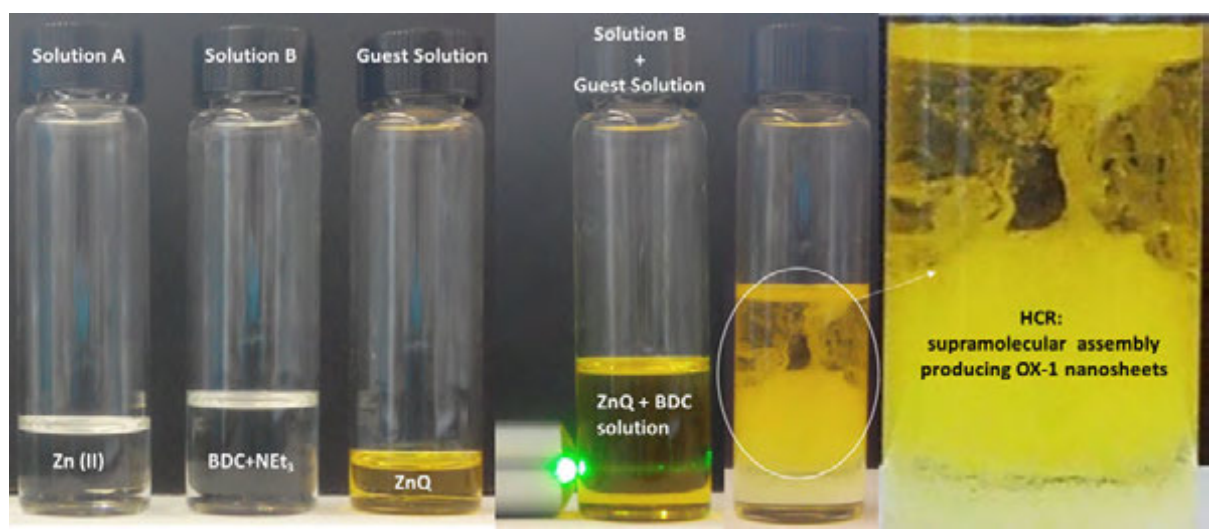

Figure shows clear solutions of Zn(II) in DMF (Solution A), BDC linker plus $NEt_3^+$ in DMF (Solution B), and ZnQ guest solution. Image on the far right shows the high-concentration reaction (HCR) when combining the A + B + ZnQ guest solutions, producing *instantaneous* supramolecular assembly containing OX-1 nanosheet materials.

To ensure that luminescent guests are not left adhered to the MOF surface, we have washed the product thoroughly twice using solvents that will solubilise the guest species (first using DMF, followed by methanol and acetone). Note that the washing step is carried out with simultaneous sonication (10 minutes per solvent, then centifugation) to further expedite the removal of external guest species. Mixture of solvents (Benzene: *N,N*-Dimethylformamide = 1:2) was used for Anthracene or Naphthalene guest solution, and the same solvent mixture was used for washing the product material to remove excess reactants.

We prepared the different "guest solutions" in accordance to the following steps:

a) **ZnQ**: ZnQ was synthesized in 1 mL of *N,N*-Dimethylformamide by reacting 1:2 molar ratio of Zn(II), (0.5 mmol) and 8-Hydroxyquinoline (8HQ), (1.0 mmol).

b) **AlQ**: AlQ was synthesized in a similar manner to that of ZnQ except by taking 1:3 molar ratio of Al(III), (0.5 mmol) and 8HQ (1.5 mmol).

c) **Naphthalene**: 0.5 mmol of Naphthalene was dissolved in 1.5 mL of Benzene and then it was mixed with 1.5 mmol solution of Zn(II) in 3 mL of *N,N*-Dimethylformamide prior to the reaction with BDC linker solution.

d) **Anthracene**: 0.5 mmol of Anthracene was dissolved in 2 mL of Benzene and then mixed with Zn(II) solution in 3mL of *N,N*-Dimethylformamide and sonicated further to prepare a clear solution.

e) **Fluorescein**: 0.1 mmol of Fluorescein was dissolved in 1 mL of DMF and mixed with Zn(II) solution before reaction with the BDC linker.

Furthermore, we have performed a systematic set of reactions to show that it is possible to control the loading of guest species confined within the OX-1 nanosheet host framework. Changes in the absorption behavior of the different products as a function of guest loading are plotted in figure below.

| Reaction # | ZnQ (Guest Concentration) | | Zn-BDC (OX-1 Concentration) | |
|---|---|---|---|---|
| | Zn(II) | 8HQ | Zn(II) | BDC$^{2-}$ |
| 1 | 0.1 mmol | 0.2 mmol | 1.5 mmol | 3.0 mmol |
| 2 | 0.3 mmol | 0.6 mmol | 1.5 mmol | 3.0 mmol |
| 3 | 0.5 mmol | 1.0 mmol | 1.5 mmol | 3.0 mmol |
| 4 | 1.0 mmol | 2.0 mmol | 1.5 mmol | 3.0 mmol |

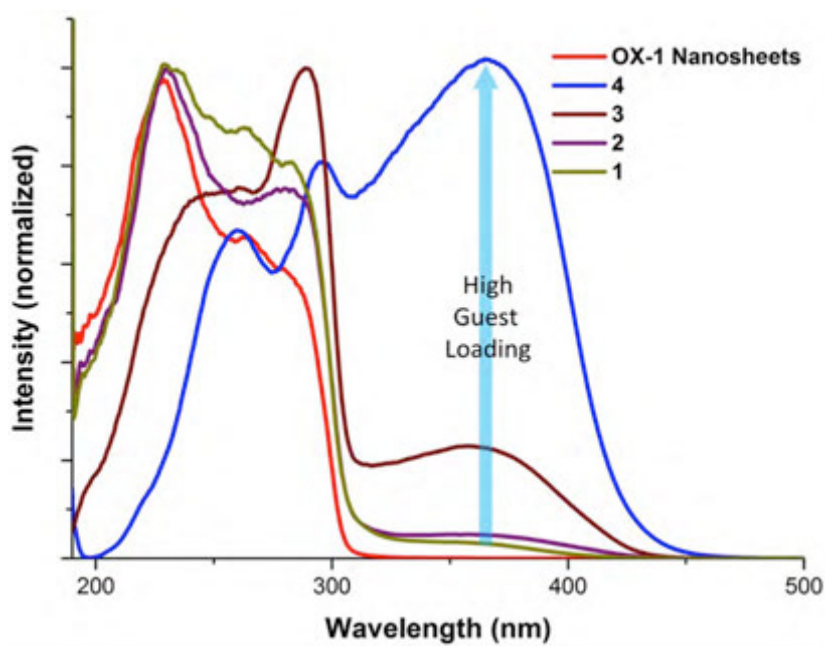

Controlled experiment with loading of various amounts of ZnQ guest emitters in OX-1 nanosheet host material (table above). Notice how the absorption spectra being modified significantly with higher guest concentration loading, signifying a stronger host-guest coupling effect.

## 2    SEM Images of Supramolecular Gels and Exfoliated Nanosheets

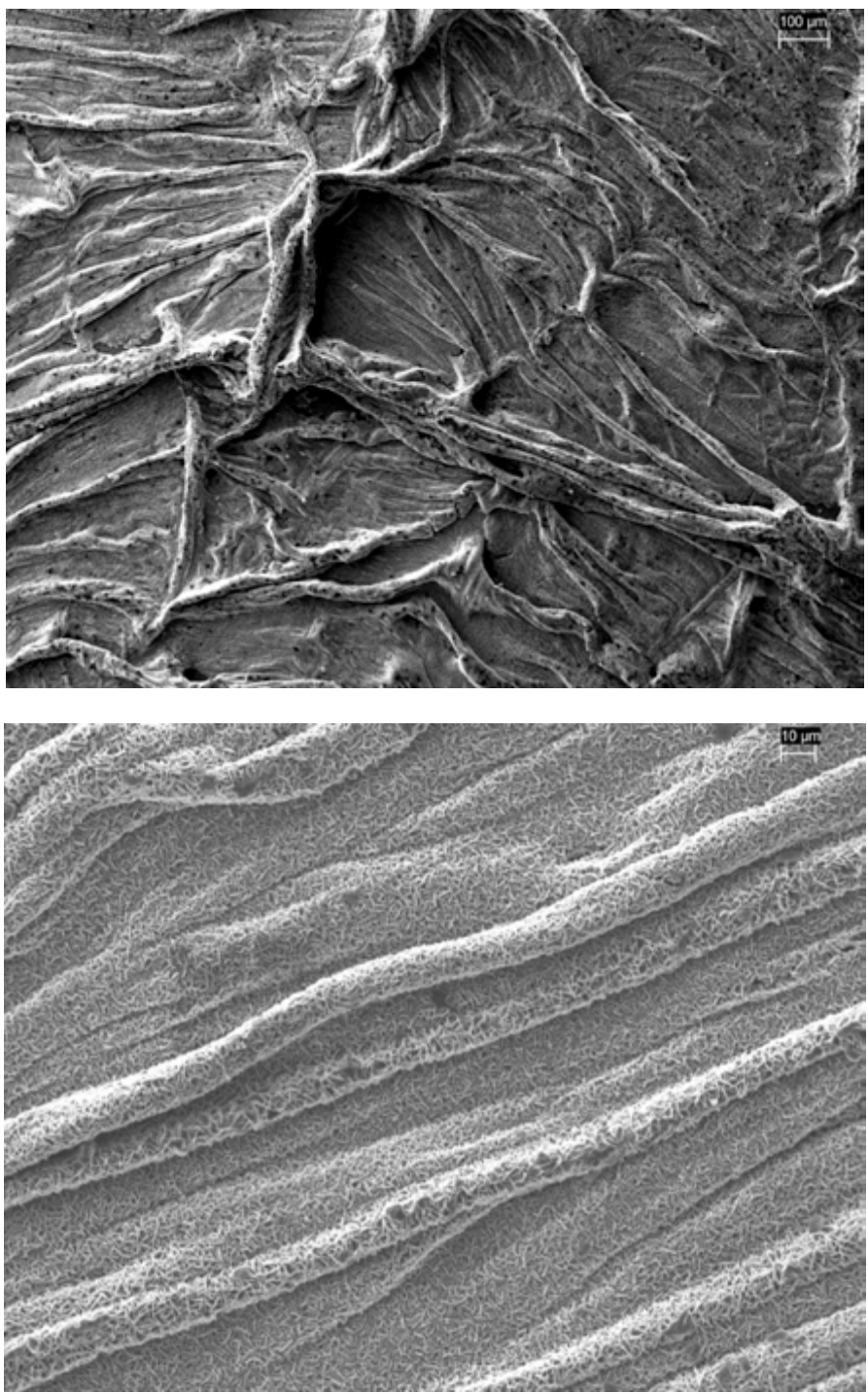

**Figure S1.** SEM images of fibrous supramolecular materials derived from high concentration reaction

(HCR), obtained by layering of reactant solutions onto a glass substrate.

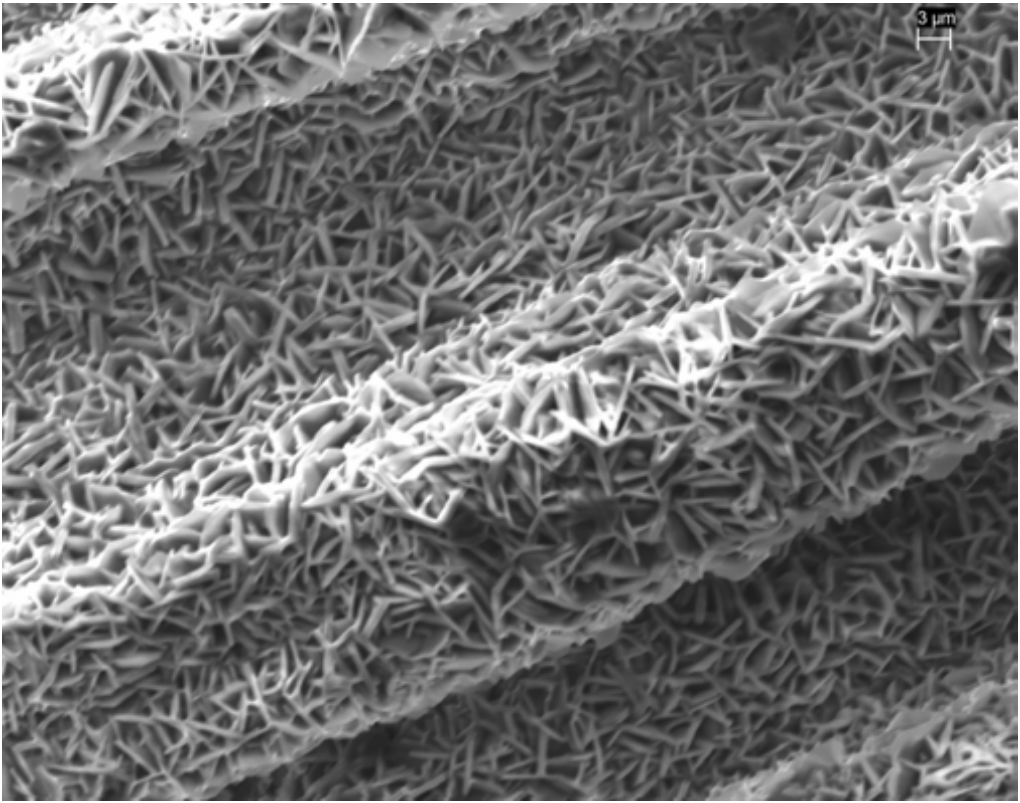

**Figure S2.** Enlarged SEM image showing densely grown 2D nanosheet structures generated through fibrous hybrid assembly at room temperature, enabled by rapid HCR approach.

## 3    TEM Images of Supramolecular Gels and Exfoliated 2D MOF Nanosheets

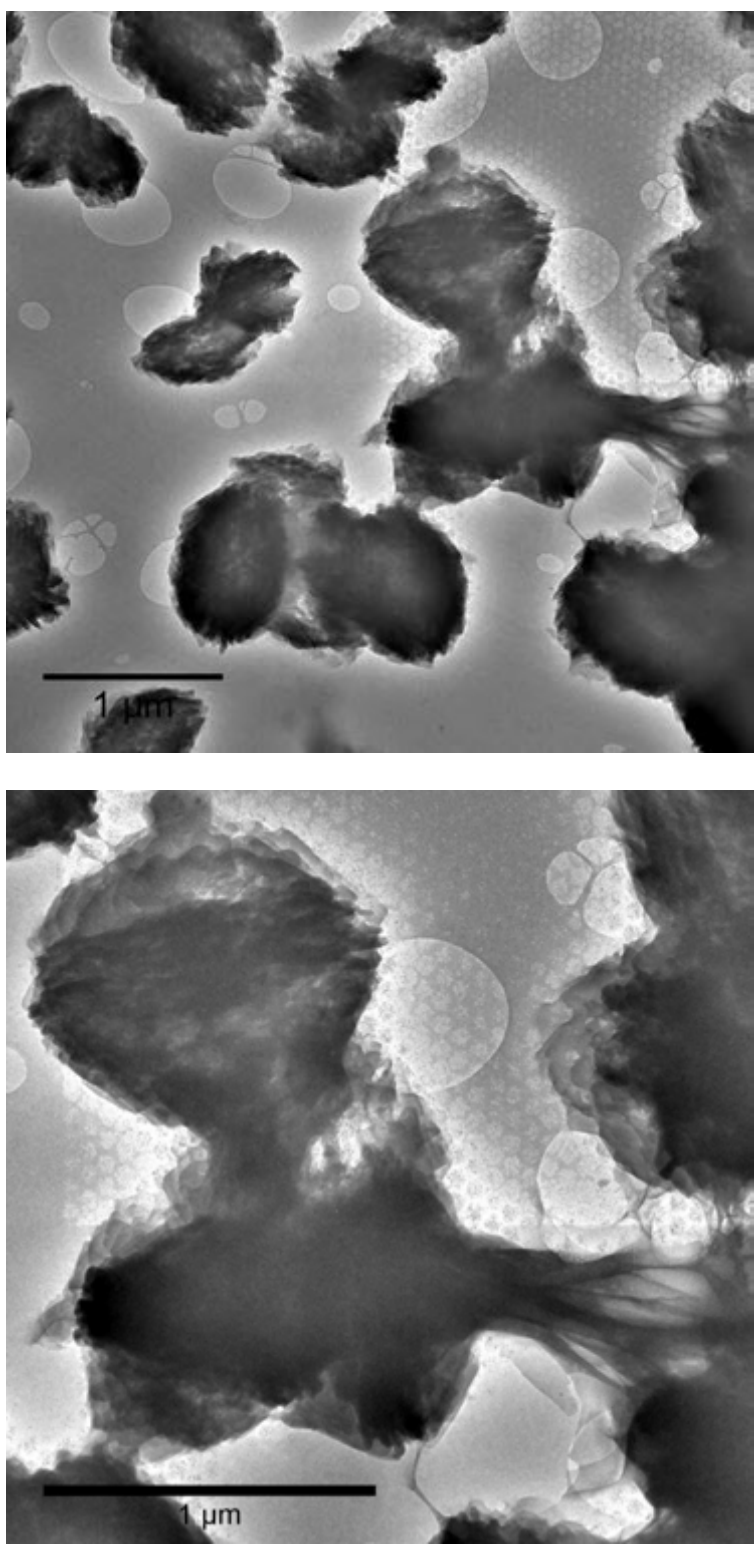

**Figure S3.** TEM image showing large clusters of 2D nanosheets embedded in a partially disintegrated supramolecular gel network. Samples are supported on a TEM grid (holey carbon background visible).

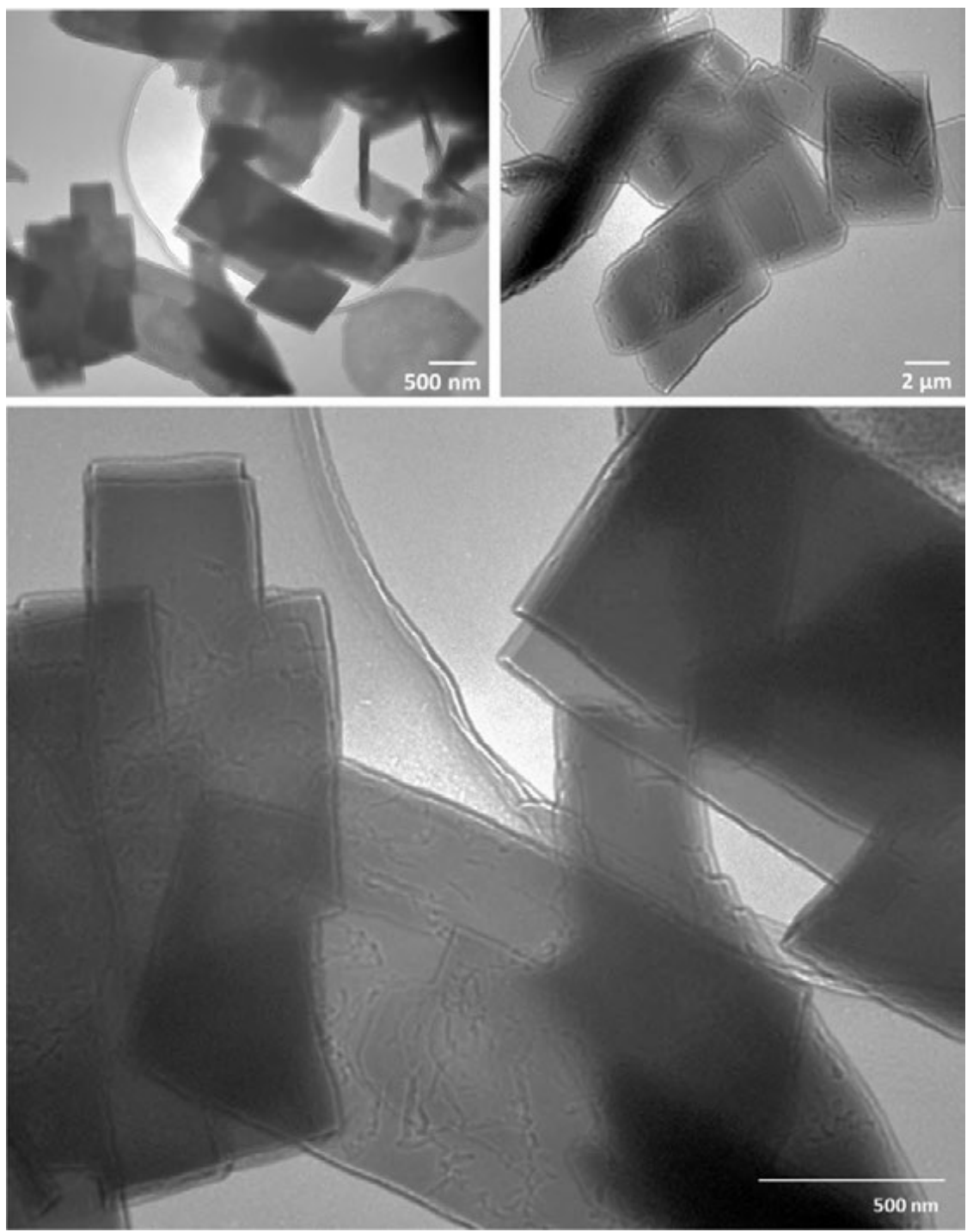

**Figure S4.** TEM showing nanosheet structures of a pristine OX-1 material revealing well defined rectangular 2D thin-sheet morphology.

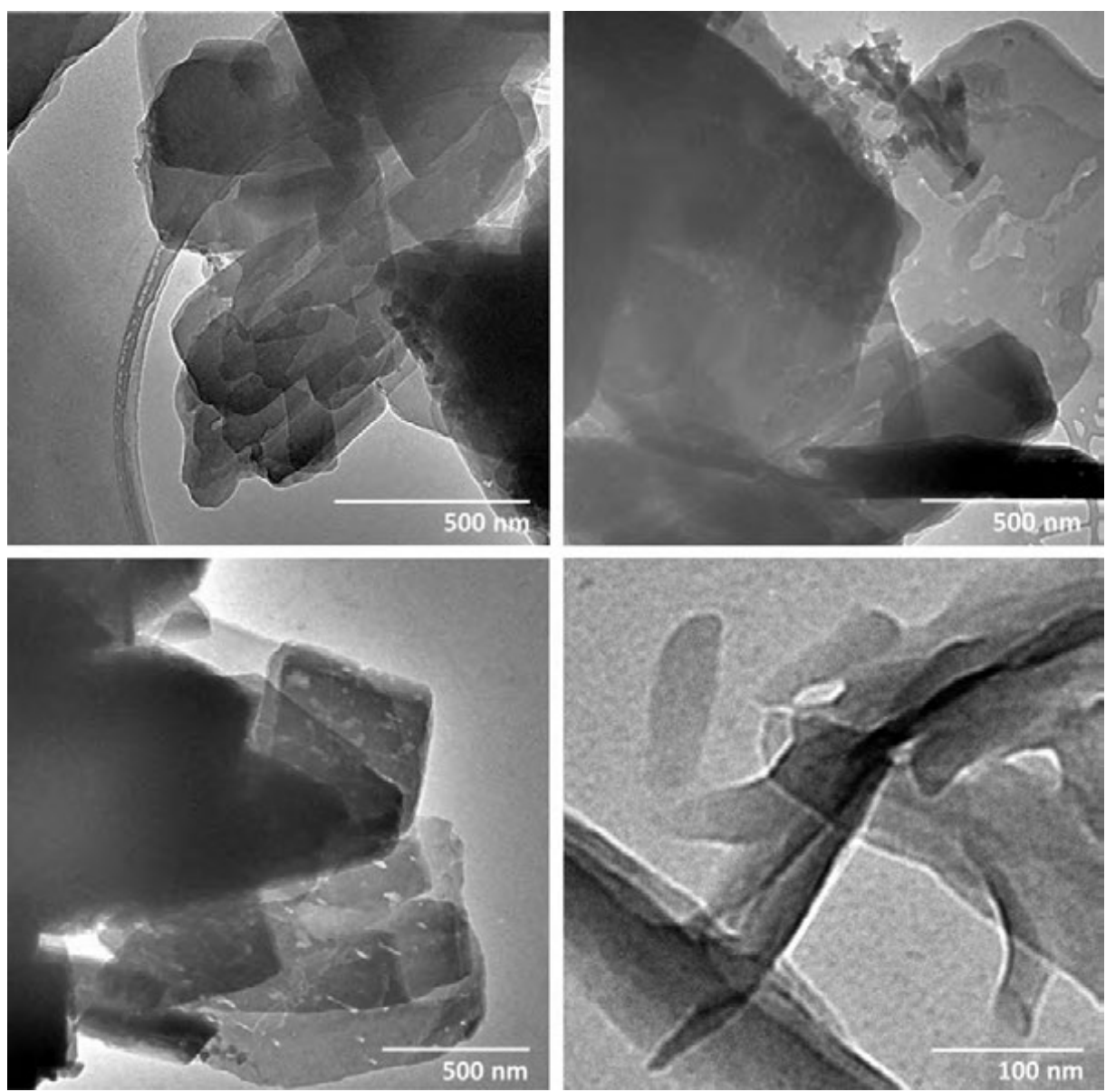

**Figure S5.** TEM micrographs of OX-1 nanosheets synthesized in DMA solvent, revealing thin 2D sheet morphology.

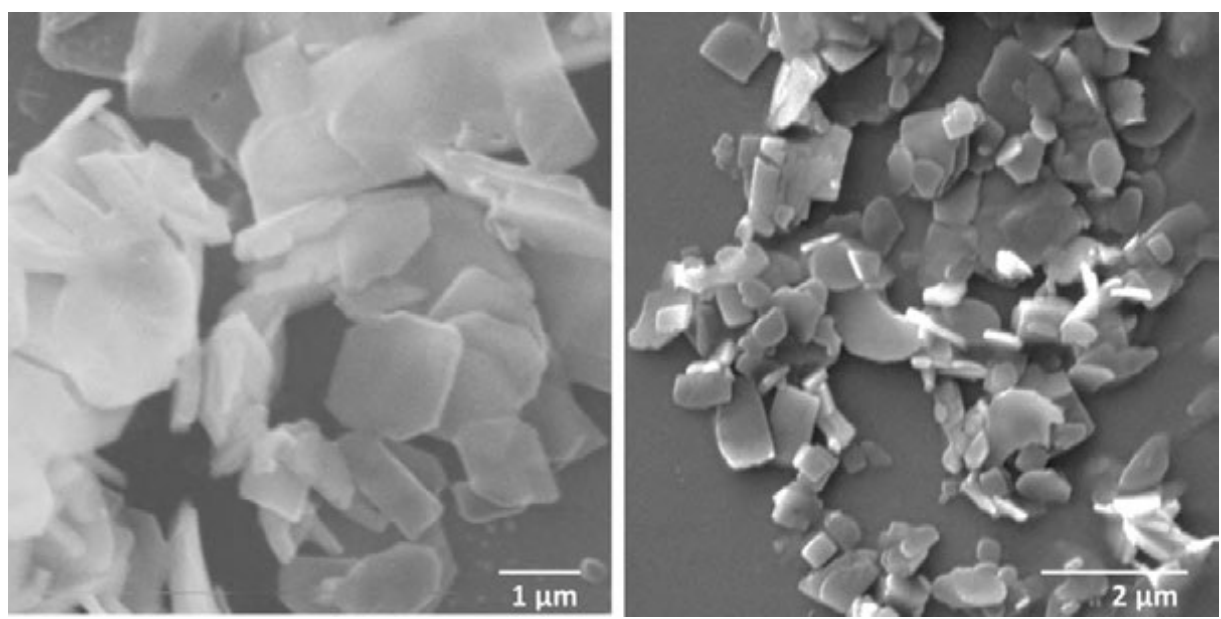

**Figure S6.** SEM micrographs of OX-1 nanosheets synthesized in DMA solvent.

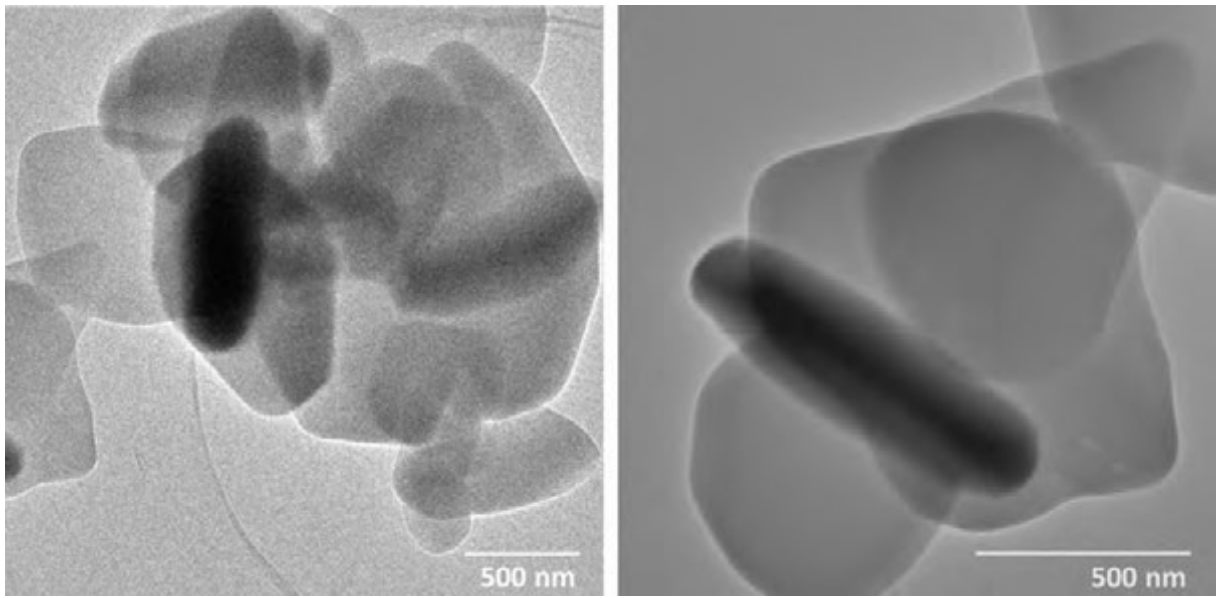

**Figure S7.** TEM micrographs of ZnQ$_{DMA}$@OX-1 nanosheets.

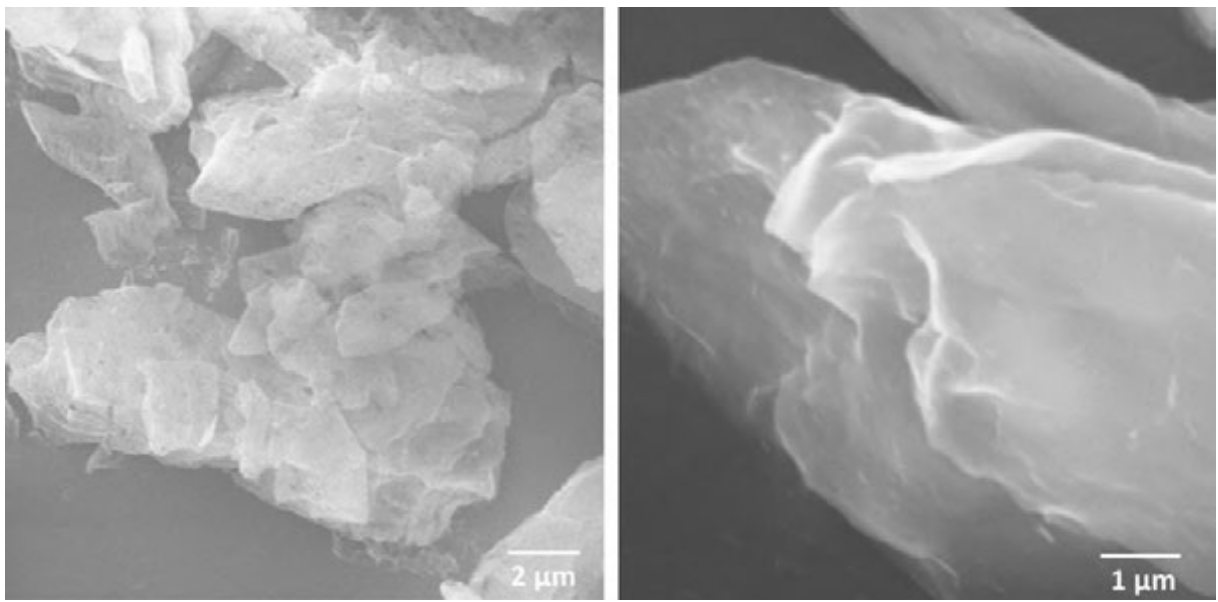

**Figure S8.** SEM micrographs of ZnQ$_{DMA}$@OX-1 nanosheets revealing thin two-dimensional morphologies.

## 4    AFM Topography Images of OX-1 Nanosheets

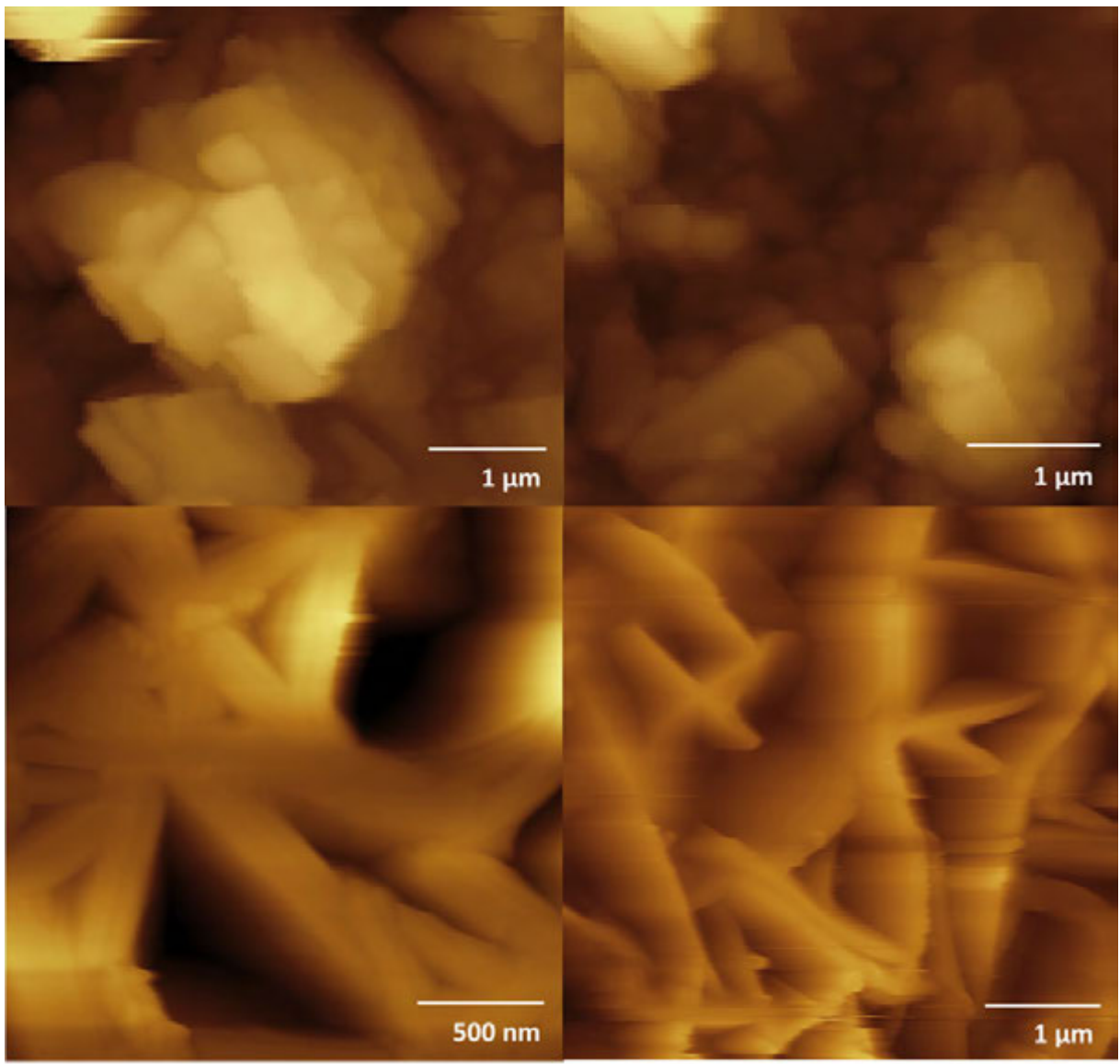

**Figure S9.** AFM showing 2D nanosheet structures and exfoliated layered morphologies of the OX-1 material.

## 5    Nanosheet Thickness Characterization

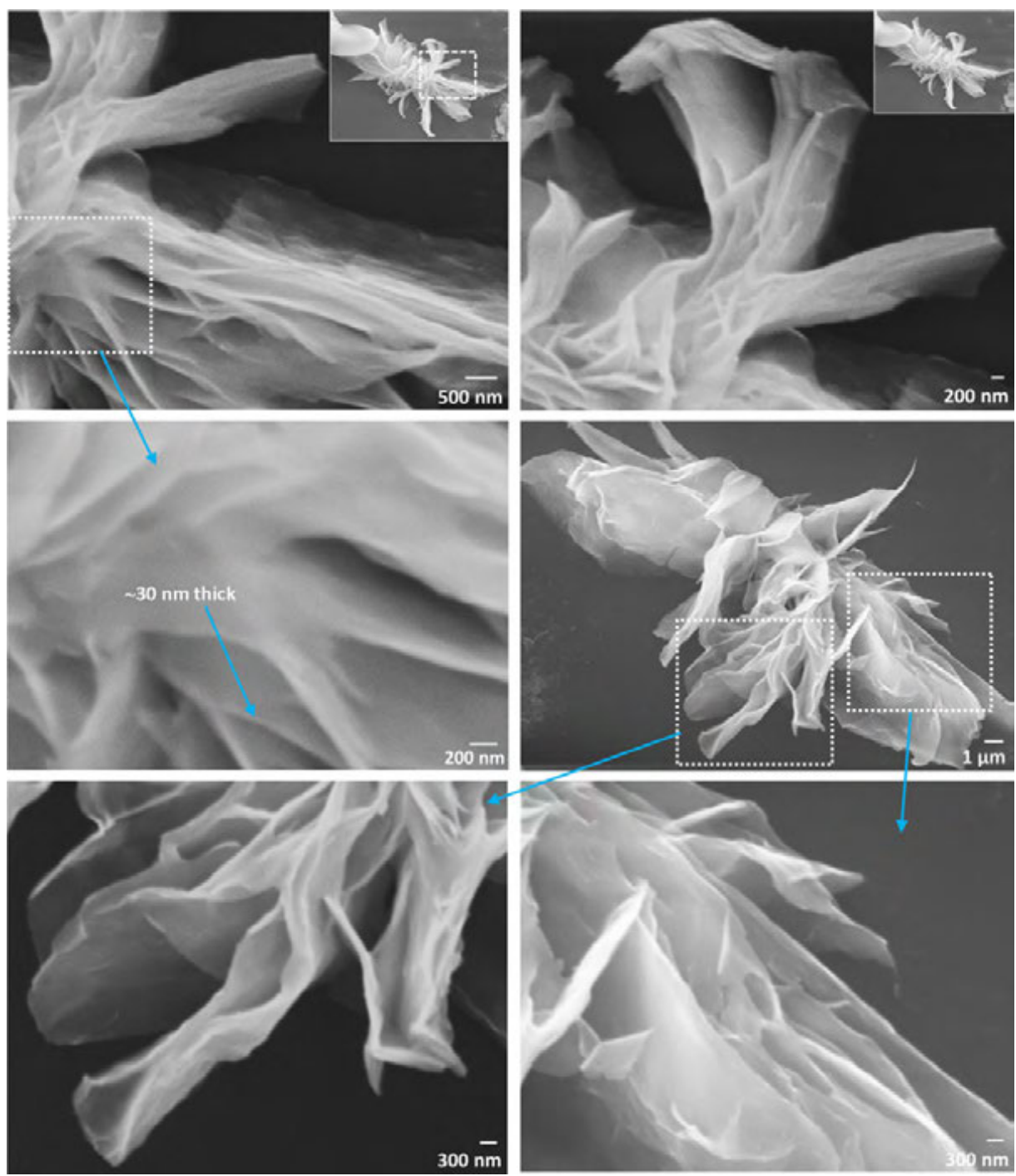

**Figure S10.** SEM imaging on functionalized nanosheet material i.e. ZnQ$_{DMF}$@OX-1 which clearly shows growth of thin extended sheets intertwined with each other as a result of high concentration reaction (HCR) approach. Layered ZnQ$_{DMF}$@OX-1 material started to delaminate after thorough washing and 1 hr sonication. These images also give insights into how multiple layers of nanosheets are stacked together prior to the exfoliation step.

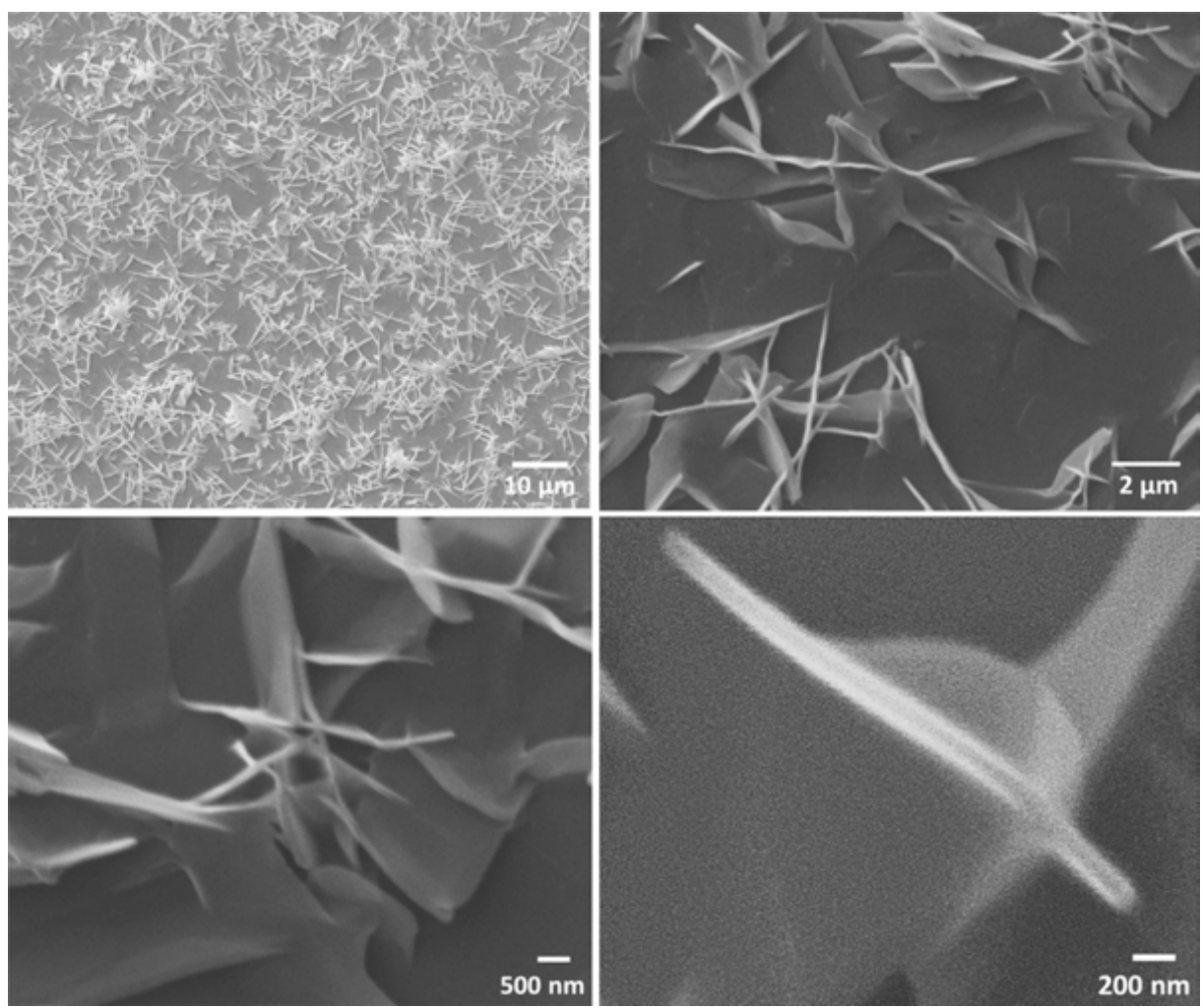

**Figure S11.** SEM of a thoroughly washed pristine OX-1 MOF material after 1 hr sonication. Prevalence of thin foils confirms facile exfoliation of OX-1 into 2D nanosheets.

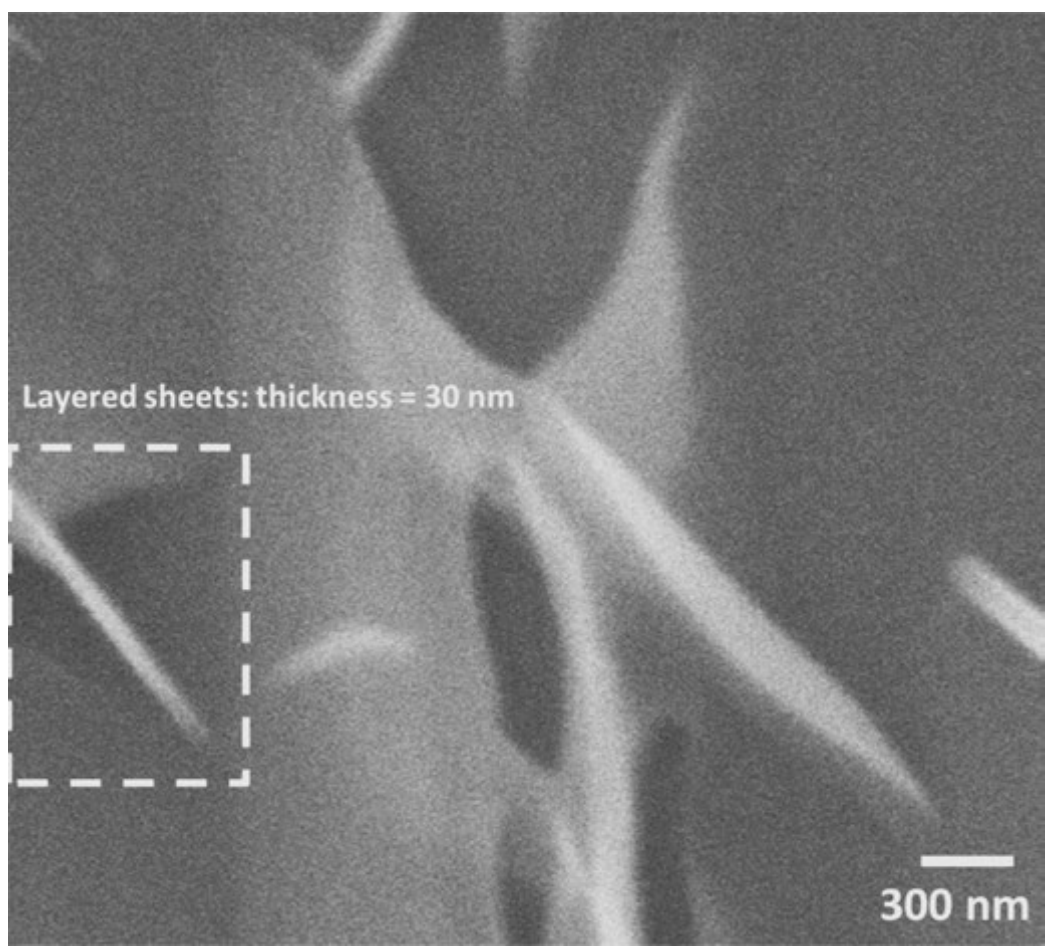

**Figure S12.** SEM of OX-1 nanosheet structures, revealing a thickness of ~30 nm of an exfoliated MOF layer.

# 6 Powder X-Ray Diffraction (PXRD) to Determine Crystal Structure of the OX-1 MOF Nanosheets

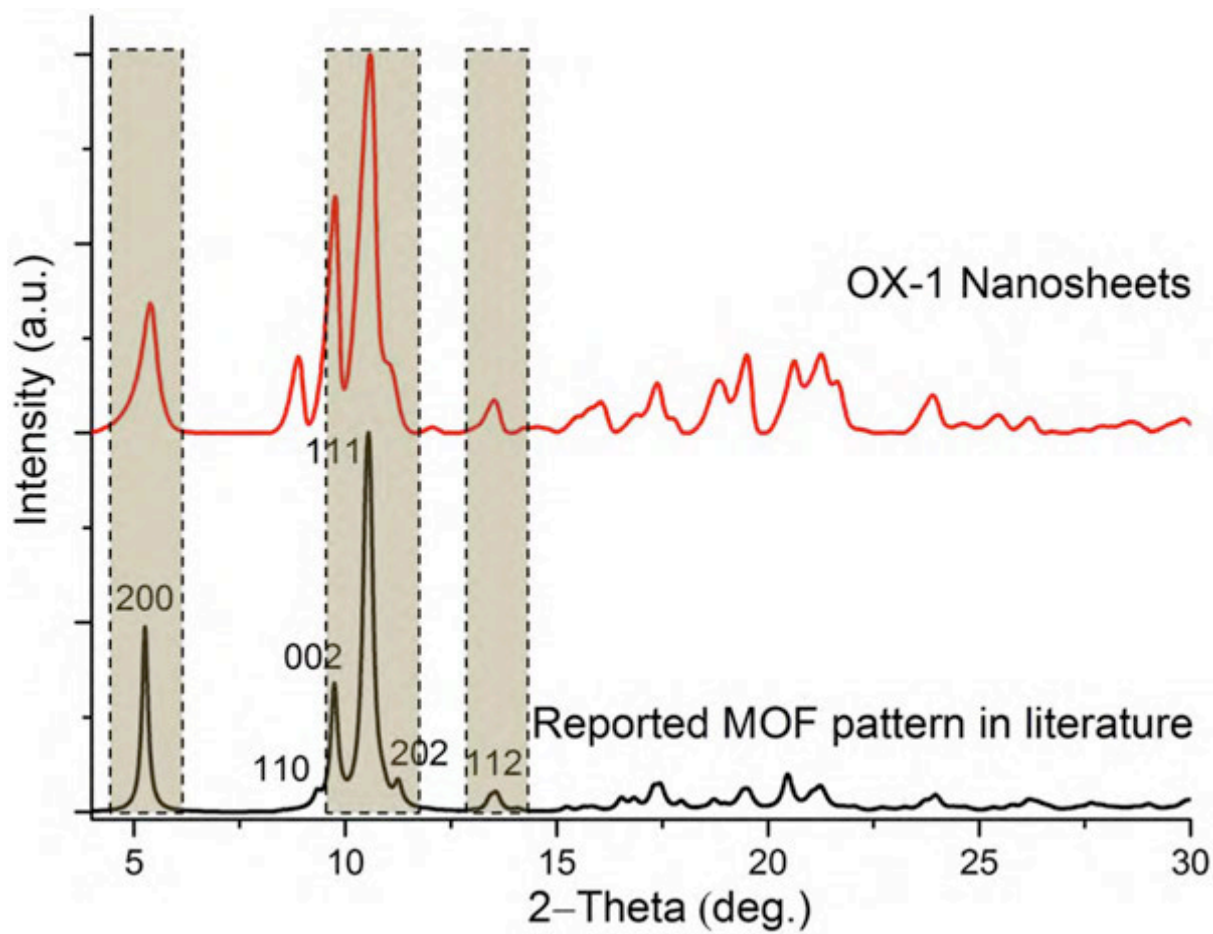

**Figure S13.** Comparison between (simulated) X-ray diffraction pattern of reported 3D MOF structure [1,2] and that of the 2D nanosheet structure experimentally obtained in current work. The detected changes in PXRD pattern are linked to lattice distortion caused by charge balancing NEt$_3^+$ cations occupying the OX-1 MOF pores. Peak broadening is associated with the nanoscale morphology of the layered OX-1 nanosheets (see TEM and SEM images).

**Table S1.** Structural parameters determined by Pawley refinement of the X-ray diffraction data (performed in Reflex module in Accelrys Material Studio v.8), compared with reported MOF structures with a higher symmetry.

| Compound | OX-1 Nanosheets | MOF reported by Burrows *et al [1]* | MOF reported by Stock *et al [2]* |
|---|---|---|---|
| Crystal System | Triclinic | Monoclinic | Monoclinic |
| Space Group | P1 | C2/c | C2/c |
| X-ray Source | CuKα | not mentioned | MoKα |
| $\lambda$ [Å] | 1.5418 | - | 0.71073 |
| Refined Range [°] | 4 to 40 | - | - |
| a [Å] | 33.576 | 33.6075 | 33.3724 |
| b [Å] | 9.8147 | 9.8727 | 9.8317 |
| c [Å] | 18.2668 | 18.1642 | 18.1967 |
| $\alpha$ [°] | 90.1257 | 90.00 | 90.00 |
| $\beta$ [°] | 90.5996 | 92.226 | 92.4550 |
| $\gamma$ [°] | 90.4146 | 90.00 | 90.00 |
| Volume [Å$^3$] | 6019.11 | 6022.27 | 5964.99 |
| Step Size | 0.02 | - | - |
| Peak Profile | Pseudo-Voigt | - | - |
| R factors | 0.084 ($R_{wp}$) | 0.0510 (R1) | 0.0395 ($R_1$) |
|  | 0.09 ($R_p$) | 0.1458 (wR2) | 0.0843 (w$R_2$) |

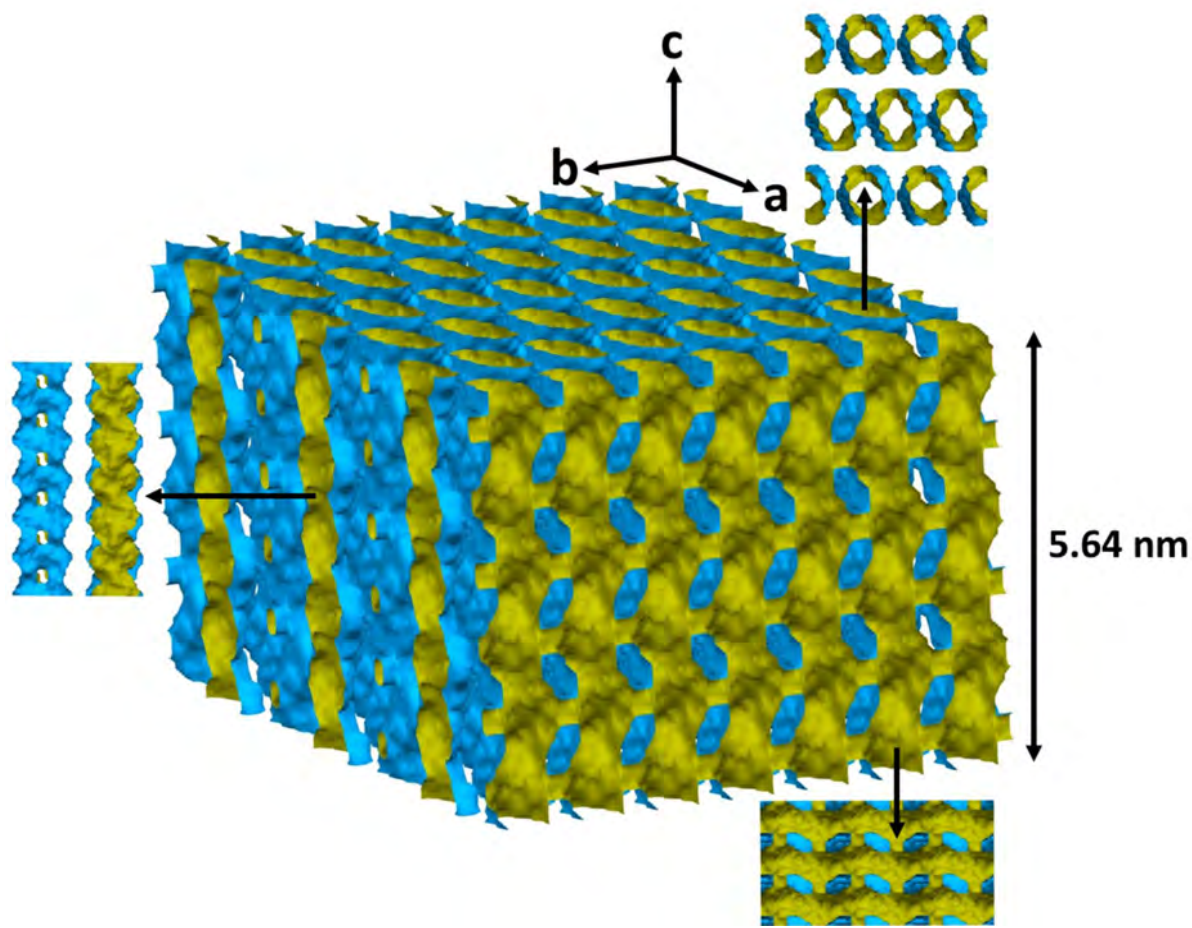

**Figure S14.** Molecular packing diagram of OX-1 MOF material showing its solvent accessible volume (SAV), which are voids that can be infiltrated by guest molecules or to be occupied by specific analytes. Undulating 1D porous channels, whose diameter is ~1 nm, are running along the *c*-axis (shown here vertically). Void volume estimated using Mercury CSD[3] was found to be 34.3% corresponding to 2062.72 Å$^3$ of empty void space per unit cell.

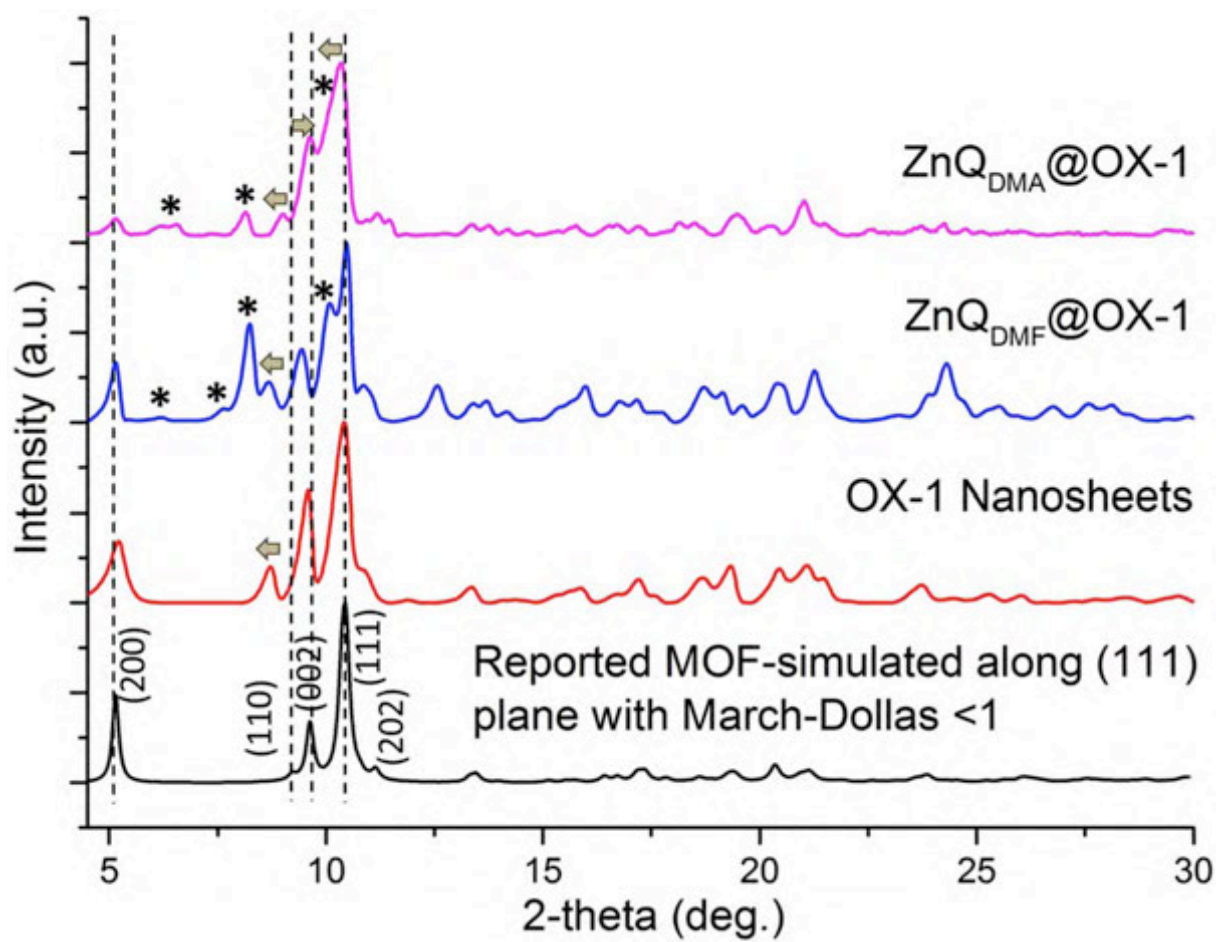

**Figure S15.** XRD patterns of MOF nanosheets with and without emissive ZnQ functional guests. Extra peaks are ascribed to the diffraction of ZnQ guest designated by the * symbol. Peak shift in OX-1 host in the case of ZnQ$_{DMA}$@OX-1 (marked by arrows on top of each peak at small angles) implies guest@host structural modification causing slight expansion in the cell dimensions.

## 7    Thermogravimetric Analysis (TGA)

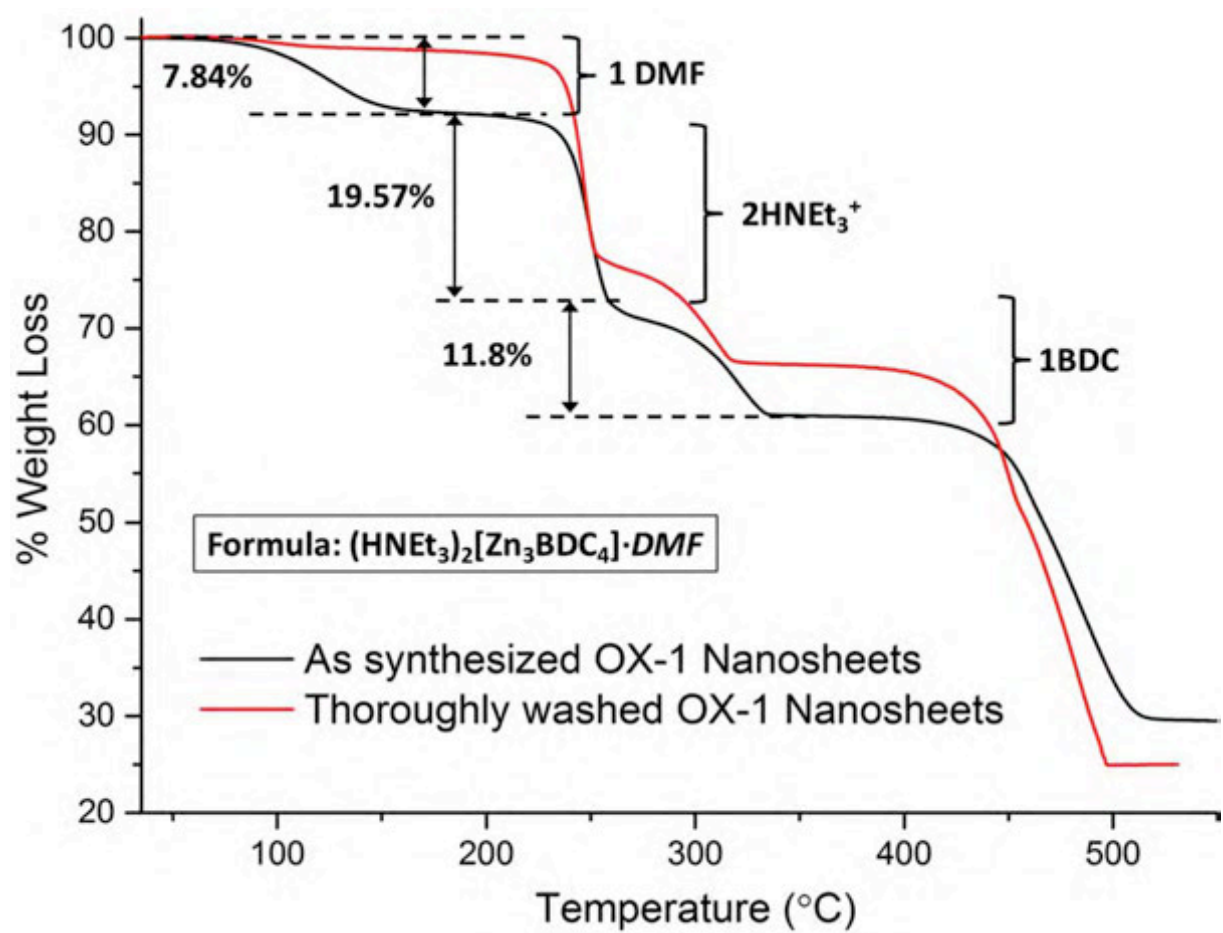

**Figure S16.** Thermogravimetric analysis (TGA) for OX-1·DMF nanosheet material synthesized in N,N-Dimethylformamide (DMF) revealing the formula for the material which is in close agreement to the expected weight loss of the components present in the material. Free solvent species occluded in the voids can be easily removed by thorough washing by low-boiling point solvent like Acetone and evacuated by heating and vacuum treatment. This behaviour is very similar to the reported MOF structure [1] where they claimed removal of BDC linker achieved in respective temperature range.

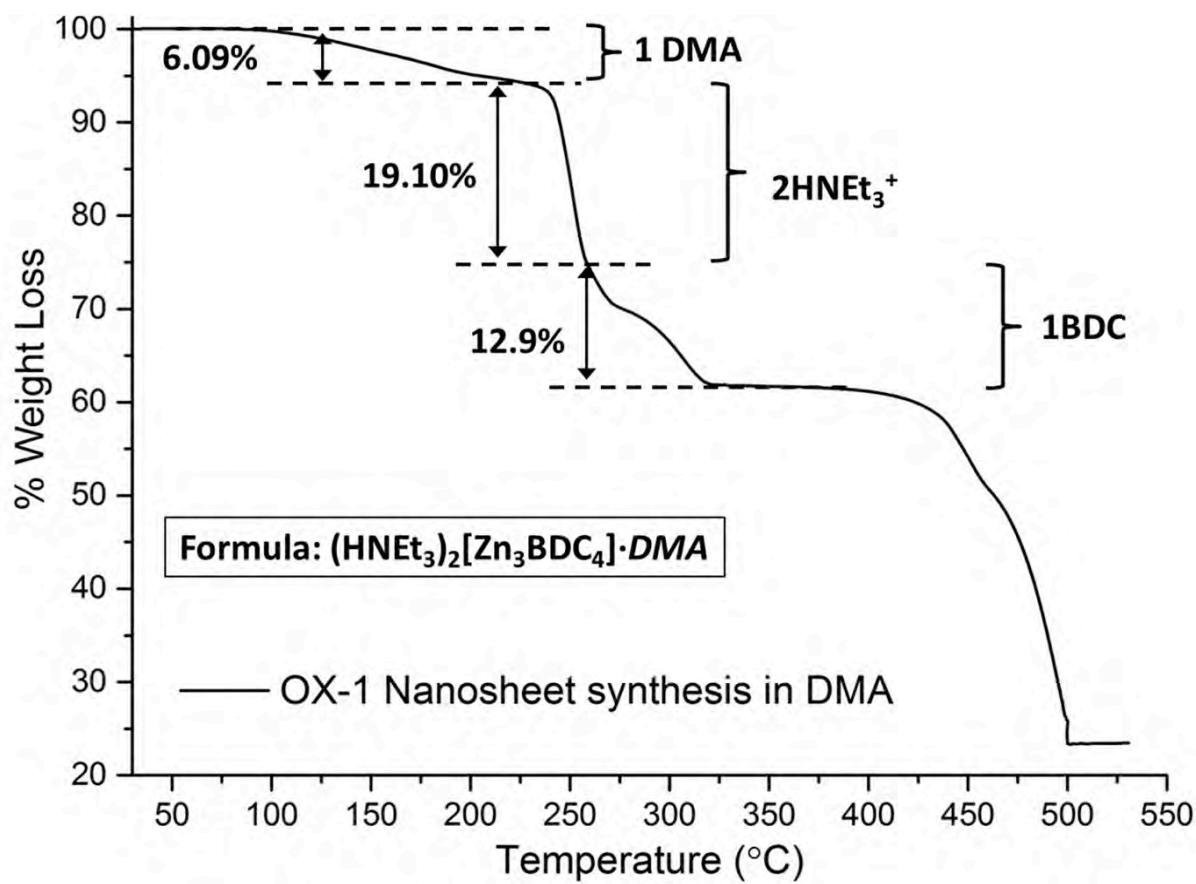

**Figure S17.** TGA for OX-1·DMA compound synthesized in N,N-Dimethylacetamide (DMA) with thorough washing and heat-vacuum treatment afterwards. Initial weight loss attributed to free solvent species that reveals entrapped solvent species cannot be removed by washing with low boiling solvent and heat-vacuum pre-treatments. (Washing step and pre-treatment conditions used were exactly the same for both compounds i.e. OX-1·DMF and OX-1·DMA, however, negligible initial weight loss for OX-1·DMF suggests removal of free solvents after pre-treatment. Conversely free solvents stay intact in OX-1·DMA even after pre-treatment).

## 8    Raman Spectroscopy

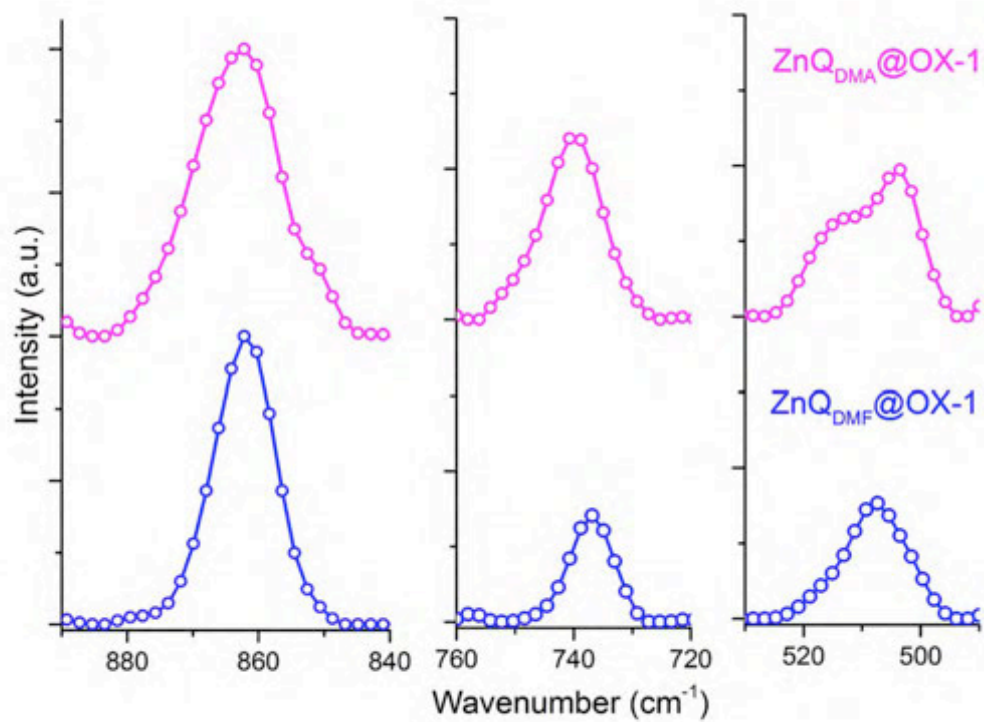

**Figure S18.** Raman modes of ZnQ$_{DMF}$@OX-1 and ZnQ$_{DMA}$@OX-1, the latter showing the doubly-degenerate bands at 852.2 cm$^{-1}$ and 514.04 cm$^{-1}$ arising from symmetry alterations of neat ZnQ guest emitter as affected by pore confinement. The full spectral range is given in main manuscript Fig. S2.

## 9 Quantum Yield and Lifetime of Pristine and Functionalized OX-1 Nanosheets

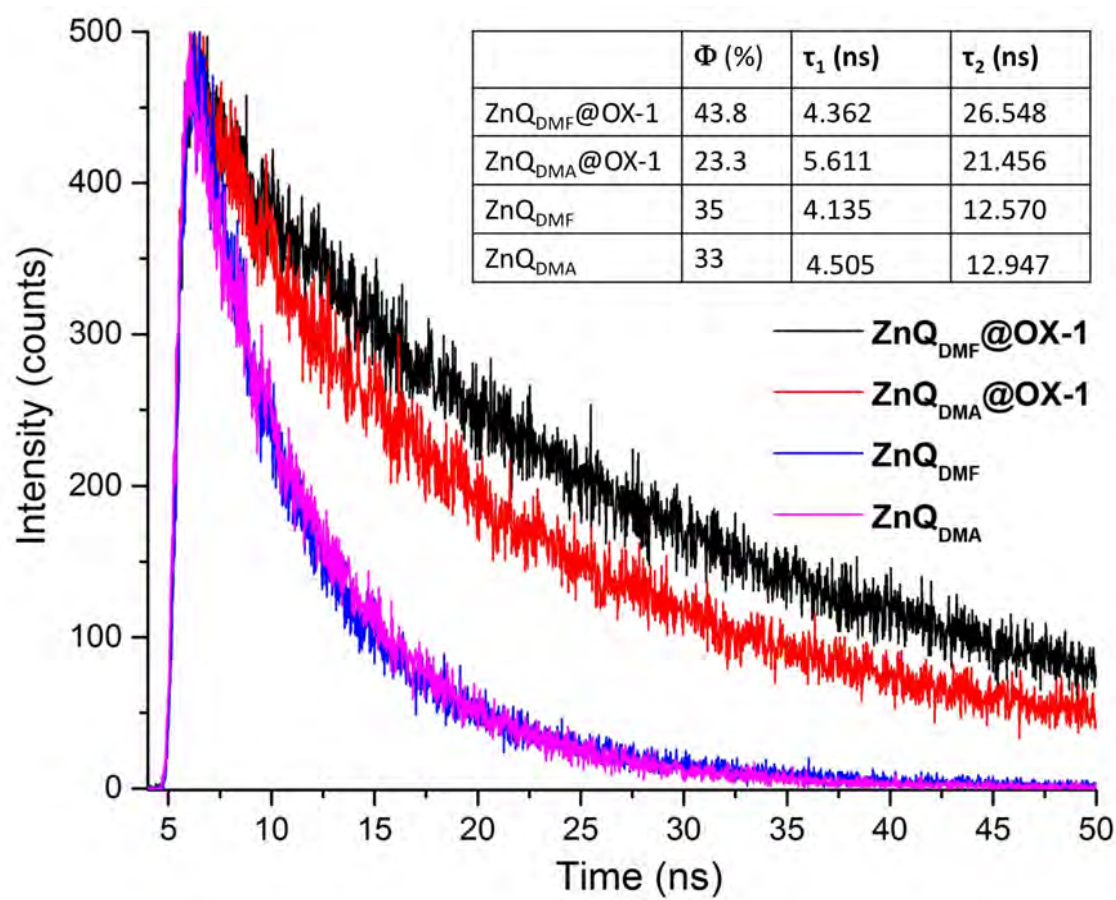

|  | Φ (%) | $\tau_1$ (ns) | $\tau_2$ (ns) |
|---|---|---|---|
| ZnQ$_{DMF}$@OX-1 | 43.8 | 4.362 | 26.548 |
| ZnQ$_{DMA}$@OX-1 | 23.3 | 5.611 | 21.456 |
| ZnQ$_{DMF}$ | 35 | 4.135 | 12.570 |
| ZnQ$_{DMA}$ | 33 | 4.505 | 12.947 |

**Figure S19.** Fluorescence lifetime decay profiles for the (pure) ZnQ guest in DMF or DMA suspensions, and the Guest@OX-1 nanosheets revealing the nanoconfinement effects on the excited-state lifetime of fluorescent guest species. Φ is quantum yield (%) and τ are time constants (ns).

## 10   Geometry Optimization and Band Gap Calculations

Geometrical optimization was performed using the Forcite module implemented in Materials Studio v.8. For the OX-1 host framework structure, molecular cluster comprising a single pore was taken into consideration by using the 1×2×1 supercell, to calculate the plausible guest geometry as a result of host confinement effects.

Density functional theory (DFT) band gap calculations were preformed using the DMol3 [4] module in Materials Studio v.8. We applied the generalized-gradient approximation (GGA) exchange-correlation functional by Perdew and Wang (PW91) [5] and the DNP basis set (double-numeric quality with polarization functions) [6]. Specific parameters used for HOMO-LUMO calculations of the host, guest, and guest@host assemblies are tabulated below:

---

**ZnQ$_{DMF}$**

# Task parameters
Calculate                 energy
Symmetry                  on
Max_memory                15000
File_usage                smart
Scf_density_convergence   1.000000e-006
Scf_charge_mixing         2.000000e-001
Scf_spin_mixing           5.000000e-001
Scf_diis                  6 pulay
Scf_iterations            50

# Electronic parameters
Spin_polarization         unrestricted
Charge                    0
Basis                     dnp
Pseudopotential           none
Functional                gga(p91)
Aux_density               octupole
Integration_grid          fine
Occupation                fermi
Cutoff_Global             4.4000 angstrom

---

**ZnQ$_{DMA}$**

# Task parameters
Calculate                energy
Symmetry                 on
Max_memory               15000
File_usage               smart
Scf_density_convergence  1.000000e-006
Scf_charge_mixing        2.000000e-001
Scf_spin_mixing          5.000000e-001
Scf_diis                 6 pulay
Scf_iterations           50

# Electronic parameters
Spin_polarization        unrestricted
Charge                   0
Basis                    dnp
Pseudopotential          none
Functional               gga(p91)
Aux_density              octupole
Integration_grid         fine
Occupation               fermi
Cutoff_Global            4.4000 angstrom

**ZnQ(Td)**

# Task parameters
Calculate                energy
Symmetry                 on
Max_memory               15000
File_usage               smart
Scf_density_convergence  1.000000e-006
Scf_charge_mixing        2.000000e-001
Scf_spin_mixing          5.000000e-001
Scf_diis                 6 pulay
Scf_iterations           50

# Electronic parameters
Spin_polarization        unrestricted
Charge                   0
Basis                    dnp
Pseudopotential          none
Functional               gga(p91)
Aux_density              octupole
Integration_grid         fine
Occupation               fermi
Cutoff_Global            4.4000 angstrom

**OX-1**

# Electronic parameters
Spin_polarization          unrestricted
Charge              0
Basis              dnp
Pseudopotential          none
Functional          gga(p91)
Aux_density          octupole
Integration_grid          medium
Occupation          fermi
Cutoff_Global          4.4000 angstrom
Kpoints          off

**ZnQ$_{DMF}$@OX-1**

Symmetry              on
Max_memory              15000
File_usage          smart
Scf_density_convergence      1.000000e-006
Scf_charge_mixing      2.000000e-001
Scf_spin_mixing          5.000000e-001
Scf_diis          6 pulay
Scf_iterations          50

# Electronic parameters
Spin_polarization          unrestricted
Charge              0
Basis              dnp
Pseudopotential          none
Functional          gga(p91)
Aux_density          octupole
Integration_grid          medium
Occupation          fermi
Cutoff_Global          4.4000 angstrom
Kpoints          off

**ZnQ(Td)@OX-1**

Symmetry              on
Max_memory              15000
File_usage          smart
Scf_density_convergence      1.000000e-005
Scf_charge_mixing          2.000000e-001
Scf_spin_mixing          5.000000e-001

```
Scf_diis               6 pulay
Scf_iterations          50

# Electronic parameters
Spin_polarization          unrestricted
Charge                 0
Basis                  dnp
Pseudopotential           none
Functional             gga(p91)
Aux_density              octupole
Integration_grid          medium
Occupation             fermi
Cutoff_Global            3.9000 angstrom
Kpoints                off
```

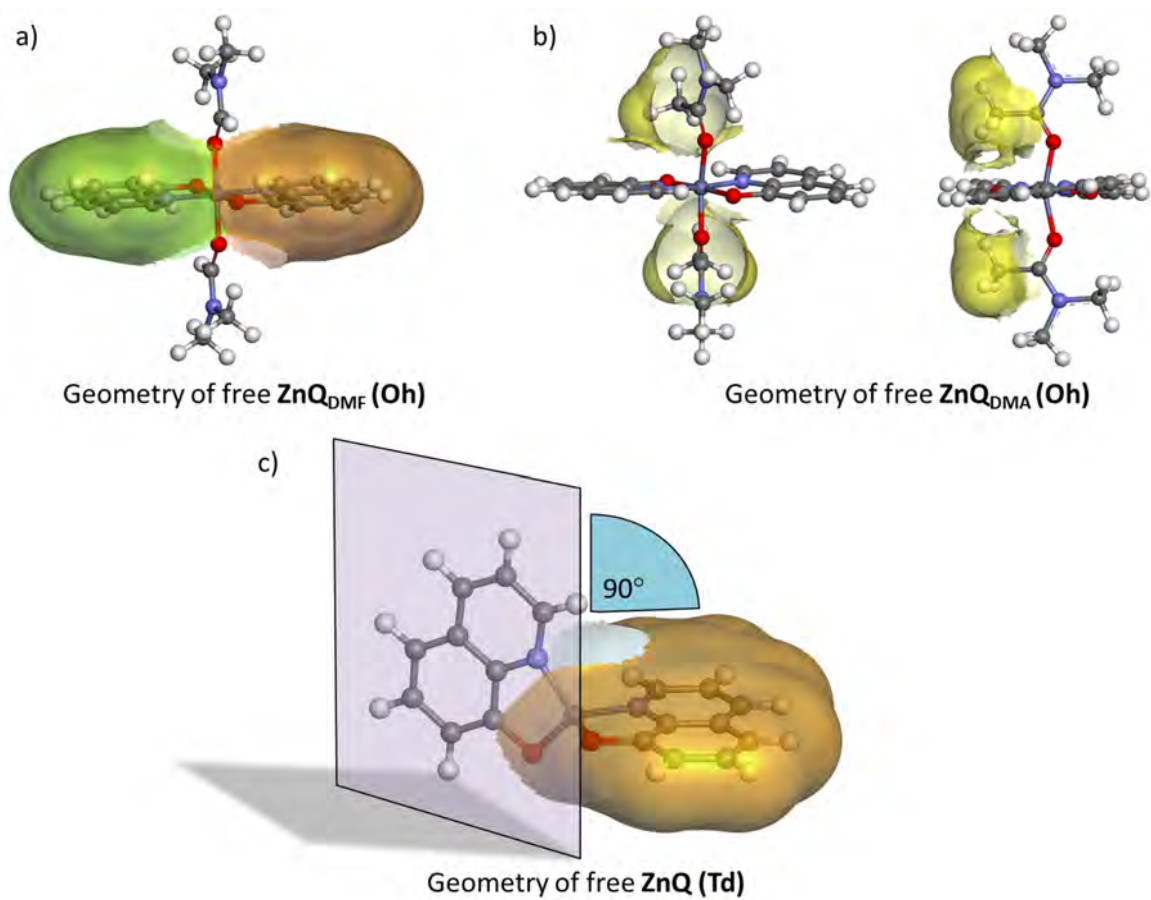

Geometry of free **ZnQ$_{DMF}$ (Oh)**

Geometry of free **ZnQ$_{DMA}$ (Oh)**

Geometry of free **ZnQ (Td)**

**Figure S20.** (a) Forcite optimized geometry of ZnQ$_{DMF}$ showing coinciding planes of two 8-hydroxyquinoline molecules coordinated to the octahedral (Oh) Zn$^{2+}$ centre, (b) ZnQ$_{DMA}$ molecule from two different views shows the non-planarity of 8-hydroxyquinoline molecule due to bulkiness of CH$_3$ groups of the DMA molecule, (c) stable configuration of the ZnQ molecule in tetrahedral (Td) geometry of Zn$^{2+}$ after removal of the two axially coordinated DMA solvent molecules.

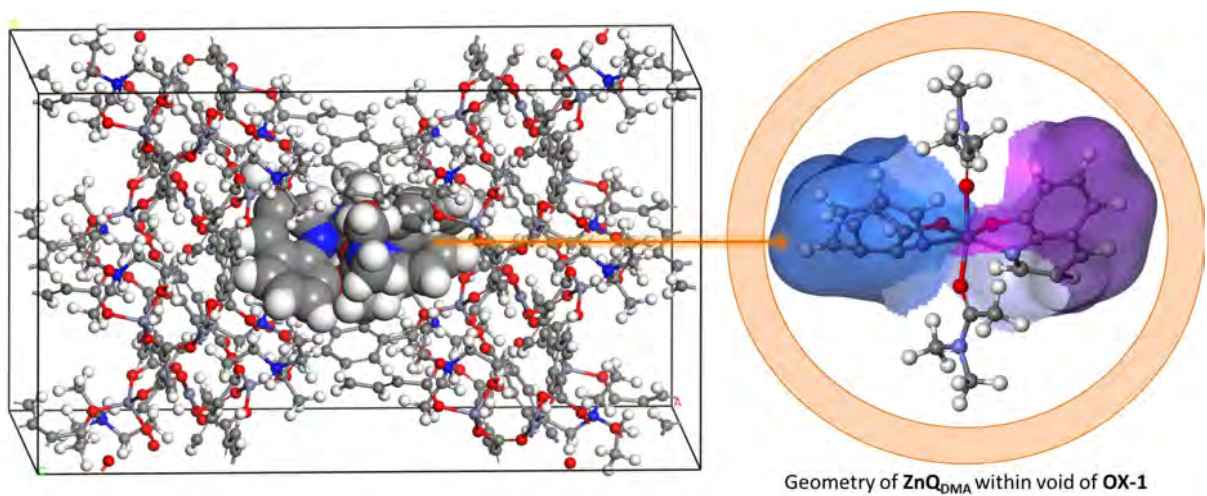

Geometry of **ZnQ$_{DMA}$** within void of **OX-1**

**Figure S21.** Extreme distortions of aromatic rings in ZnQ$_{DMA}$ molecule in confined conditions of OX-1, suggesting unfavourable configuration for in-situ guest encapsulation and plausible removal of DMA molecules from the axial positions of Zn$^{2+}$ centres during framework formation (in actual experiment).

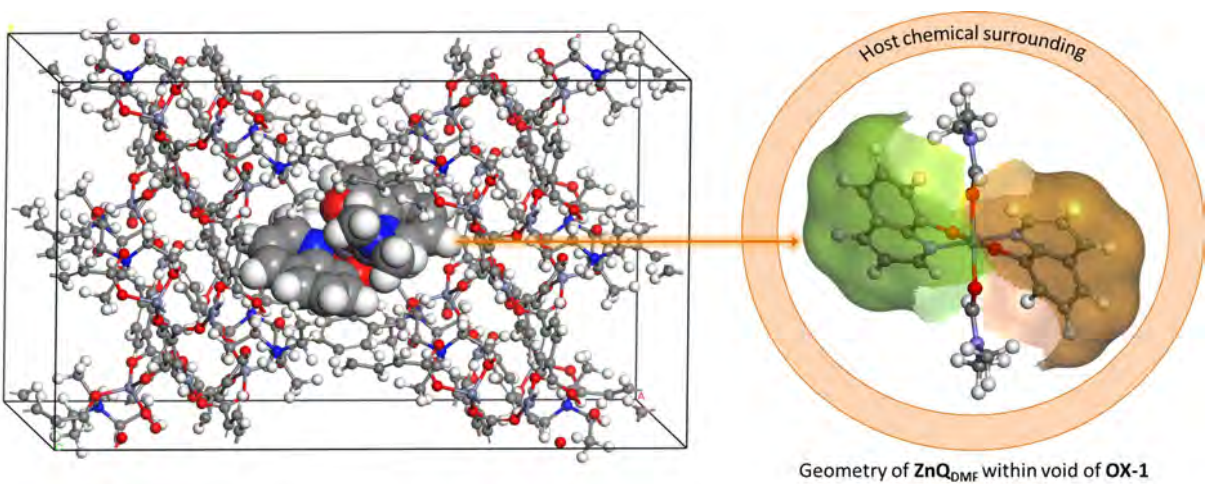

Geometry of **ZnQ$_{DMF}$** within void of **OX-1**

**Figure S22.** Optimised geometry of ZnQ$_{DMF}$ molecule when constrained within the void of OX-1. Geometry was optimised using Forcite module in Materials Studio. Figure shows flipping of aromatic rings and slight distortion in the molecule due to confinement effects, which would cause modification in band gaps and resulting optical properties.

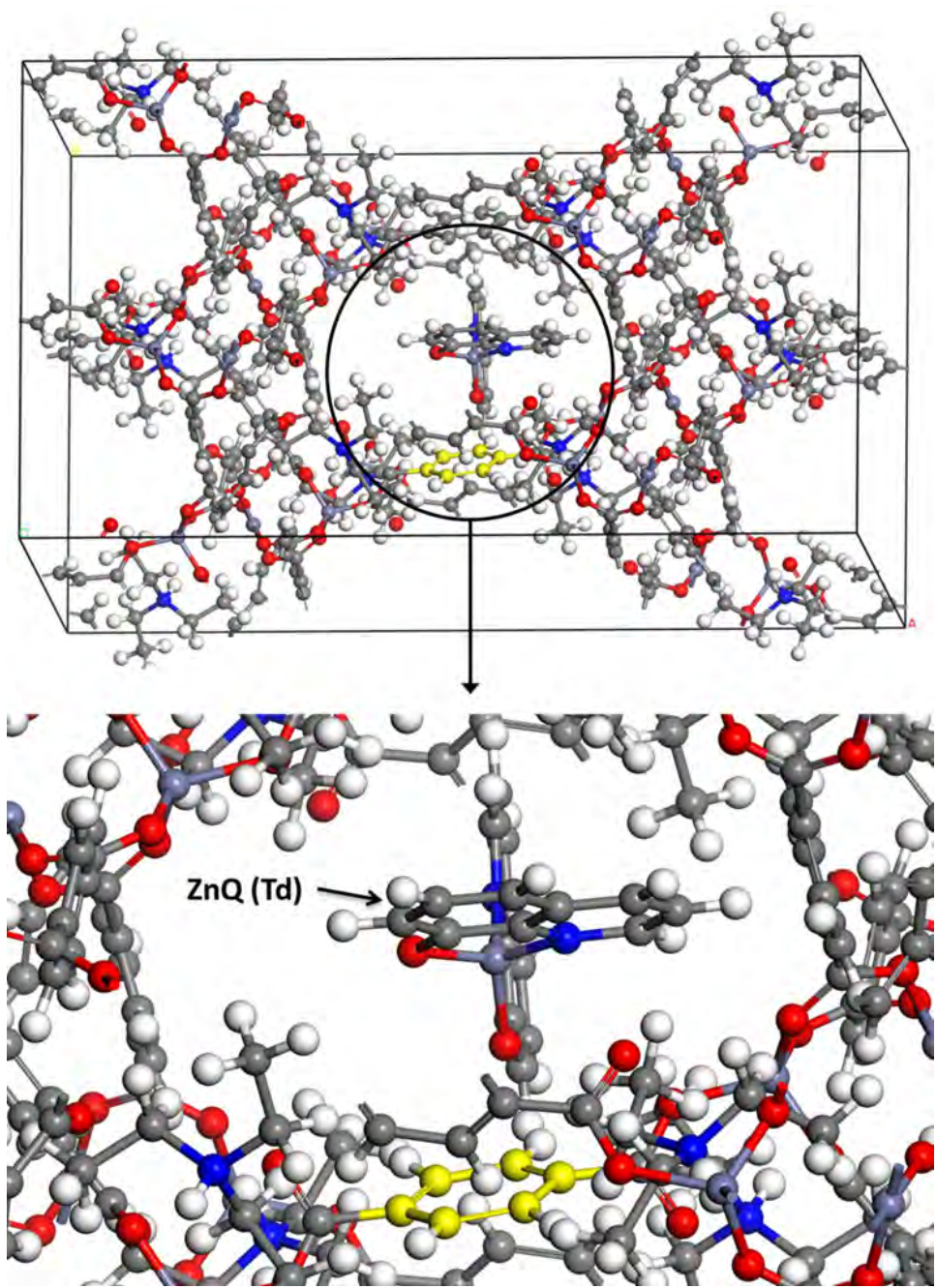

**Figure S23.** Stable geometry of ZnQ(Td) configuration (after removal of DMA molecules from the axial positions of $Zn^{2+}$ centres) making weak interactions with the aromatic linkers of the OX-1 host. This host-guest configuration supports red shift in emission properties of $ZnQ_{DMA}@OX-1$ compound (Fig.2e) due to π-π interactions and H-bonding. Yellow highlighted ring represents an aromatic benzene ring from linker of host material.

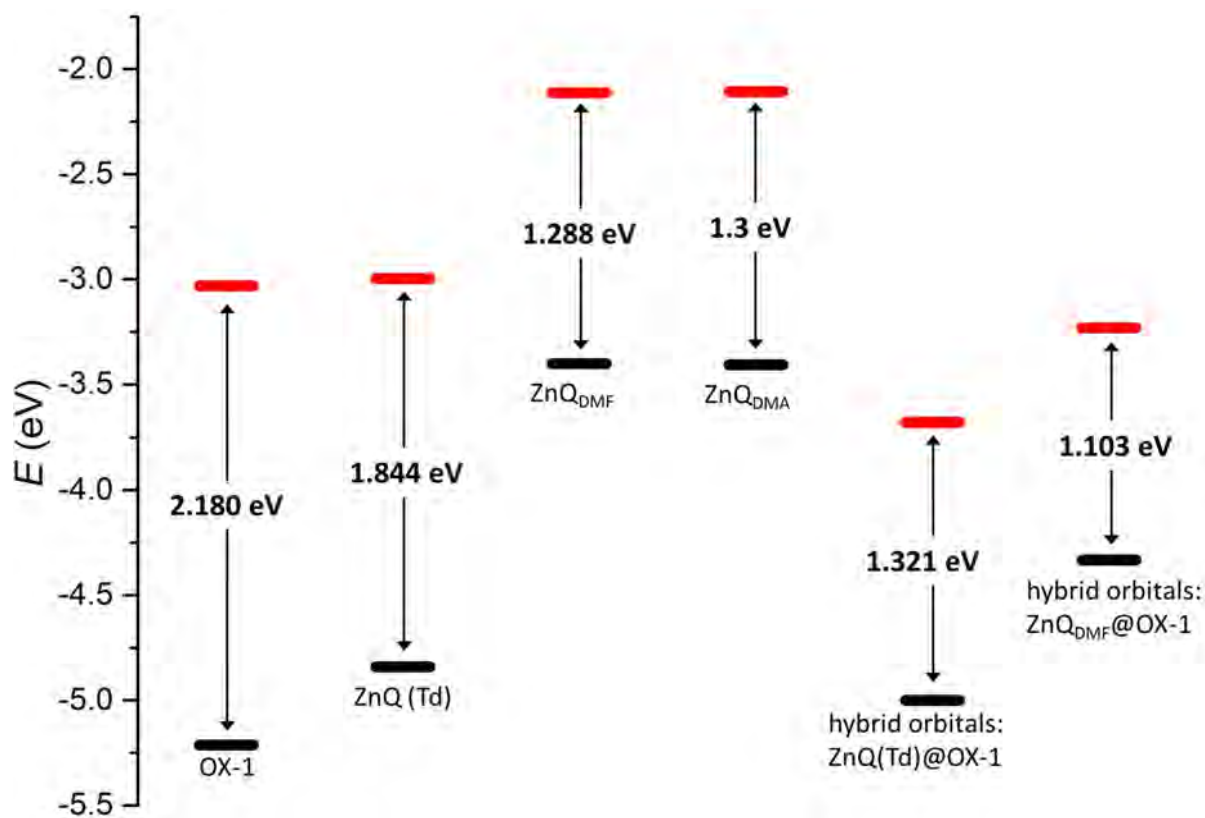

**Figure S24**. DFT calculated band gap energy values of the OX-1 host framework (1×2×1 supercell) and ZnQ guest emitter considering different possible host-guest configurations (DMF, DMA, Td). Formation of hybrid orbitals in Guest@OX-1 lowers the energy levels due to intimate host-guest interactions, shown on the right for the case of ZnQ(Td)@OX-1 and ZnQ$_{DMF}$@OX-1. Note that ZnQ$_{DMA}$@OX-1 is equivalent to ZnQ(Td)@OX-1, i.e. without DMA solvent coordination at its two axial positions (see Fig.S20b).

**Table S2.** HOMO and LUMO orbitals of different species involved in current work and their calculated band gap values (underestimation by DFT is well recognized [7] compared with experiments). Blue and yellow isosurfaces denote the positive and negative charges respectively.

| Molecule | HOMO | LUMO | Band Gap (eV) |
|---|---|---|---|
| ZnQ (Td) | 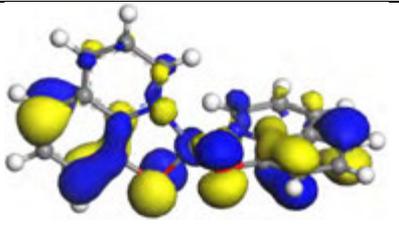 | 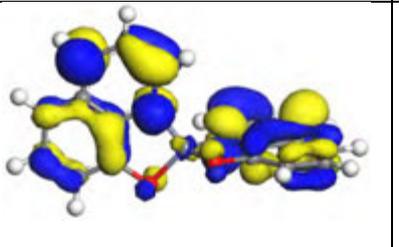 | 1.844 |
| ZnQ$_{DMF}$ | 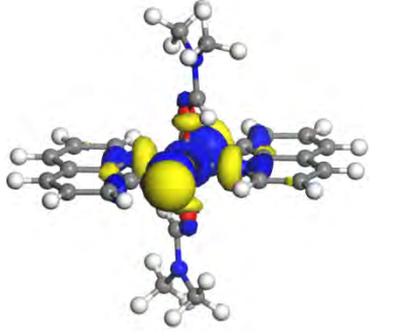 | 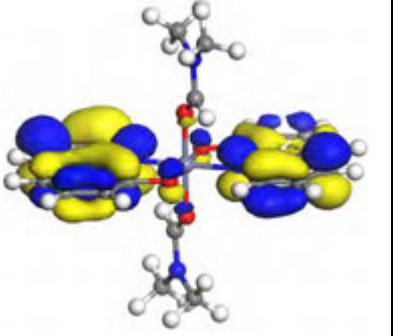 | 1.288 |
| ZnQ$_{DMA}$ | 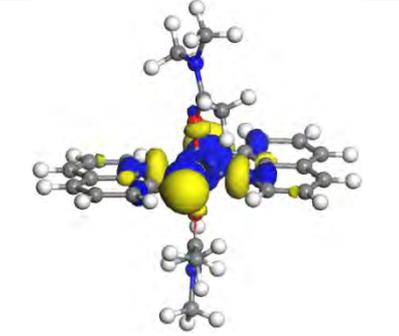 | 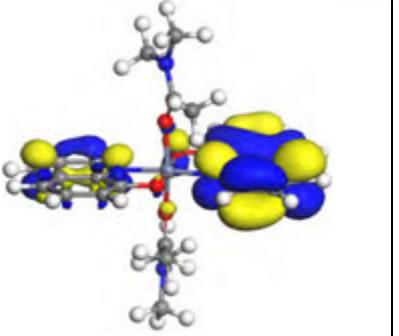 | 1.3 |
| OX-1 | 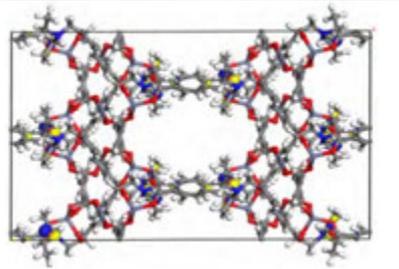 | 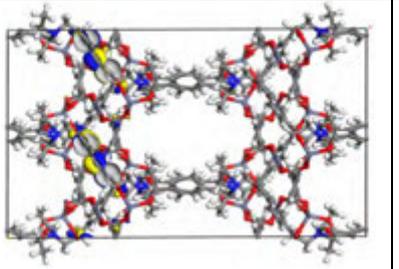 | 2.180 |
| ZnQ(Td)@OX-1 (equivalent to ZnQ$_{DMA}$@OX-1 obtained from DMA synthesis) | 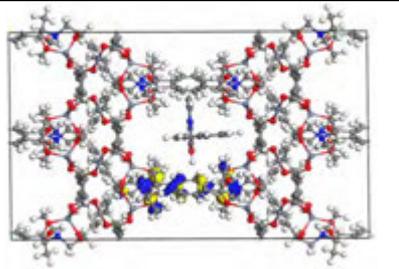 | 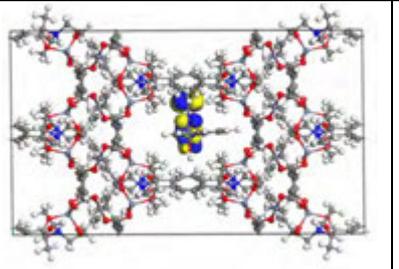 | 1.321 |

| | | |
|---|---|---|
| ZnQ$_{DMF}$@OX-1 | 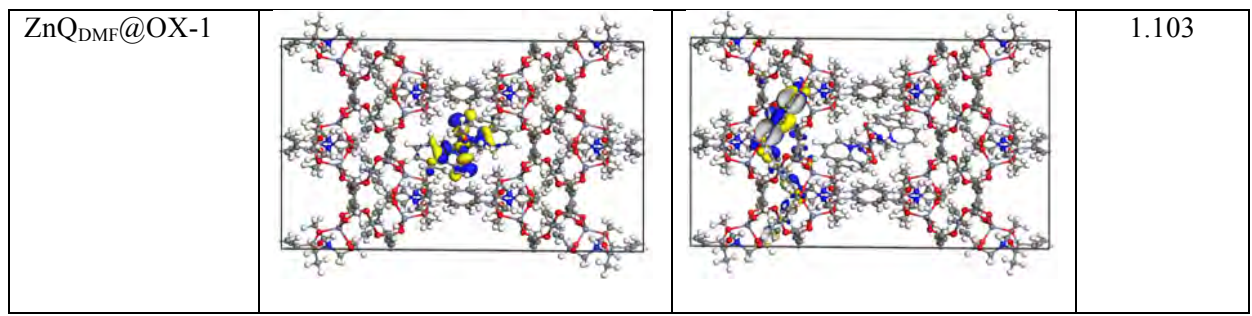 | 1.103 |

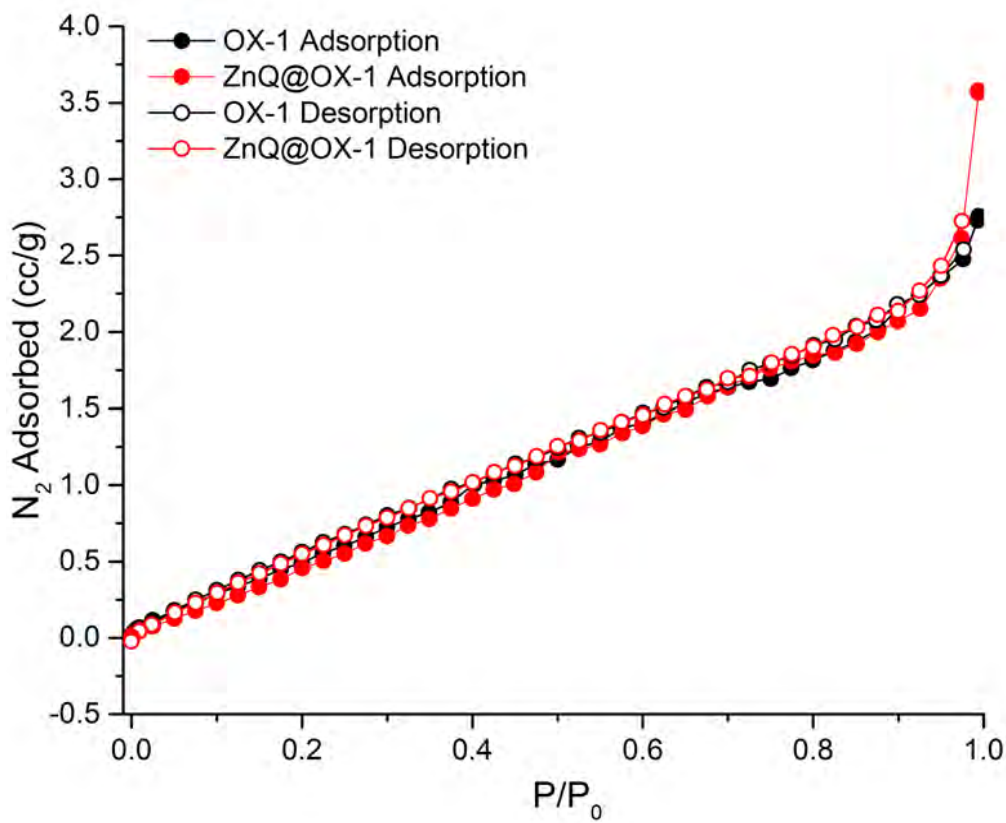

**Figure S25**. $N_2$ adsorption profile at 77 K for $ZnQ_{DMF}$@OX-1 and OX-1. Both samples were evacuated in the same way by heating at 80 °C under vacuum for 40 hr. Interestingly, we found that the isotherms of both compounds were similar, and with a relatively low BET surface area of ~80 m$^2$/g. This value is reminiscent of the surface area of a similar framework structure ((H$_2$NEt$_2$)$_2$[Zn$_3$(BDC)$_4$]·3DEF, Langmuir surface ~66 m$^2$/g) reported by Stock et al [2]. Negligible uptake suggests that OX-1 based nanosheets are non-reactive towards inert gas like N$_2$, or there is also possibility of pore blockages due to restrictions in sample activation to remove entrapped species. Above notwithstanding, we note that OX-1 has a strong chemical affinity towards VOC analyte sensing (see Fig.S26-S27).

## 11 Host-Guest Interaction and Optochemical Stimulation

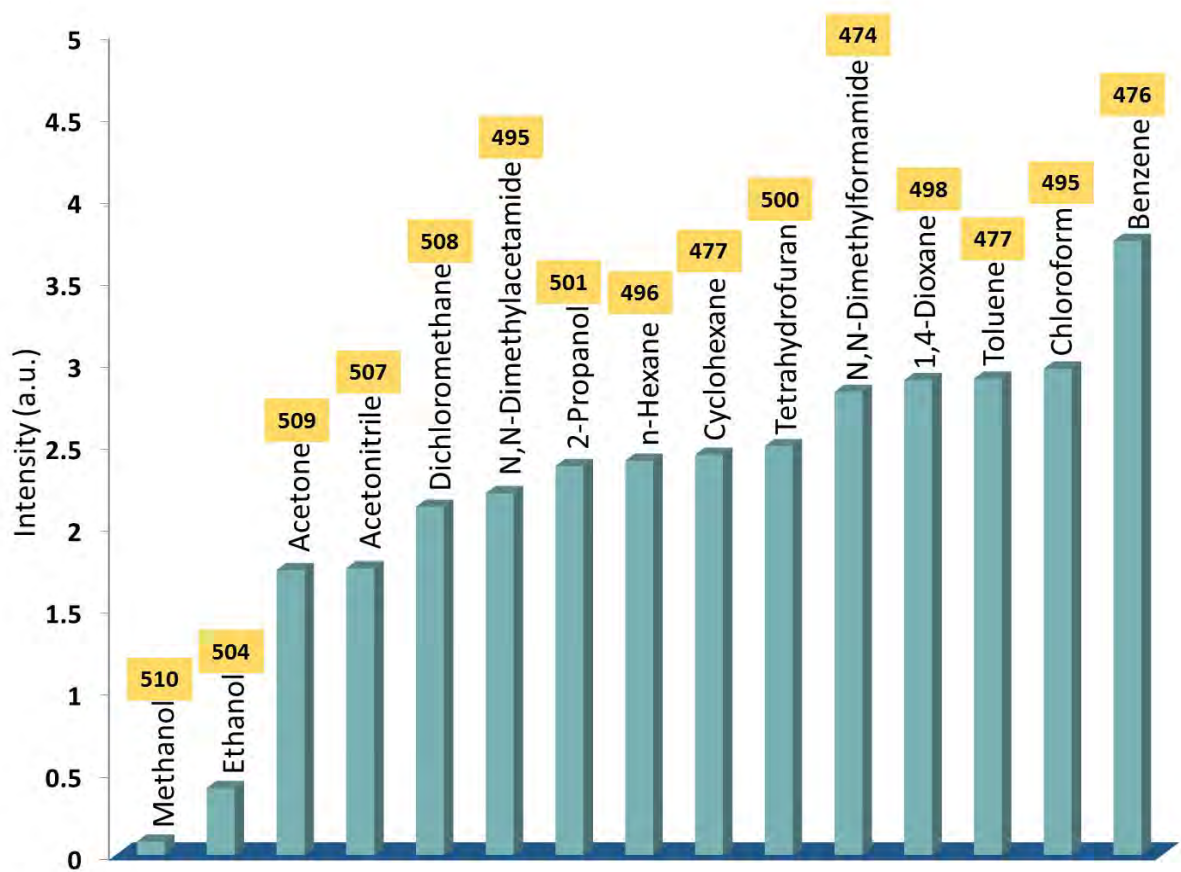

**Figure S26.** Small-molecule sensing ability of ZnQ$_{DMF}$@OX-1 nanosheets *via* changing emission intensity and shifting in wavelength due to optochemical perturbations in MOF pores. Numbers on top of each bar denotes the emission wavelength λ$_{em}$ (nm).

Solvent analyte properties tabulated with the decreasing order of luminescence intensity of functionalized ZnQ$_{DMF}$@OX-1 material after analyte interaction. Solvent accessible surface area and volume are evaluated using Accerlys Discovery Studio Visualizer. α, β, and π* are Kamlet-Taft [8] parameters describing ability of solvent to donate the proton, to accept the proton, and solvent dipolarity (polarizability), respectively.

| No. | Solvent | Volume (Å³) | Surface Area (Å²) | β | α | π* | λ$_{em}$ |
|-----|---------|-------------|-------------------|------|-------|-------|----------|
| 1 | Benzene | 80.11 | 100.793 | 0.1 | 0 | 0.59 | 476 |
| 2 | Chloroform | 72.684 | 93.789 | 0 | 0.44 | 0.58 | 495 |
| 3 | Toluene | 97.081 | 118.204 | 0.11 | 0 | 0.54 | 477 |
| 4 | Dioxane | 83.543 | 102.749 | 0.37 | 0 | 0.55 | 498 |
| 5 | DMF | 75.089 | 97.715 | 0.76 | 0 | 0.88 | 474 |
| 6 | THF | 73.885 | 93.672 | 0.55 | 0 | 0.58 | 500 |
| 7 | Cyclohexane | 98.279 | 114.683 | 0 | 0.001 | 0 | 477 |
| 8 | n-Hexane | 109.141 | 134.561 | 0 | 0 | -0.08 | 496 |
| 9 | 2-Propanol | 68.471 | 90.491 | 0.95 | 0.76 | 0.48 | 501 |
| 10 | DMA | 92.094 | 113.992 | 0.69 | 0 | 0.88 | 495 |
| 11 | DCM | 56.697 | 78.34 | 0 | 0.30 | 0.82 | 508 |
| 12 | ACN | 44.166 | 65.95 | 0.31 | 0.19 | 0.75 | 507 |
| 13 | Acetone | 62.921 | 85.788 | 0.48 | 0.08 | 0.71 | 509 |
| 14 | Ethanol | 51.117 | 73.915 | 0.77 | 0.83 | 0.54 | 504 |
| 15 | Methanol | 34.54 | 54.982 | 0.62 | 0.93 | 0.6 | 510 |

Decrease in luminescence

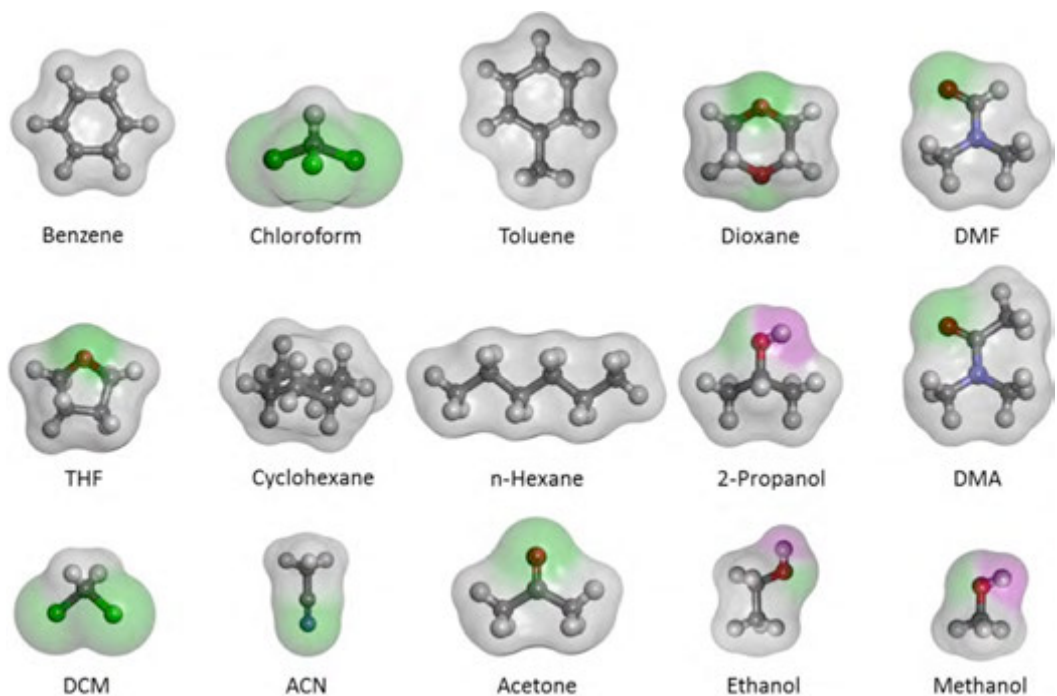

**Figure S27.** Hydrogen bond donor and acceptor surfaces of solvent analytes indicating hydrophobic sites in gray, hydrogen bond acceptor site in green, and hydrogen bond donor site in red color. Table above shows surface area and volume in numbers tabulated in the order of change in luminescence intensity.

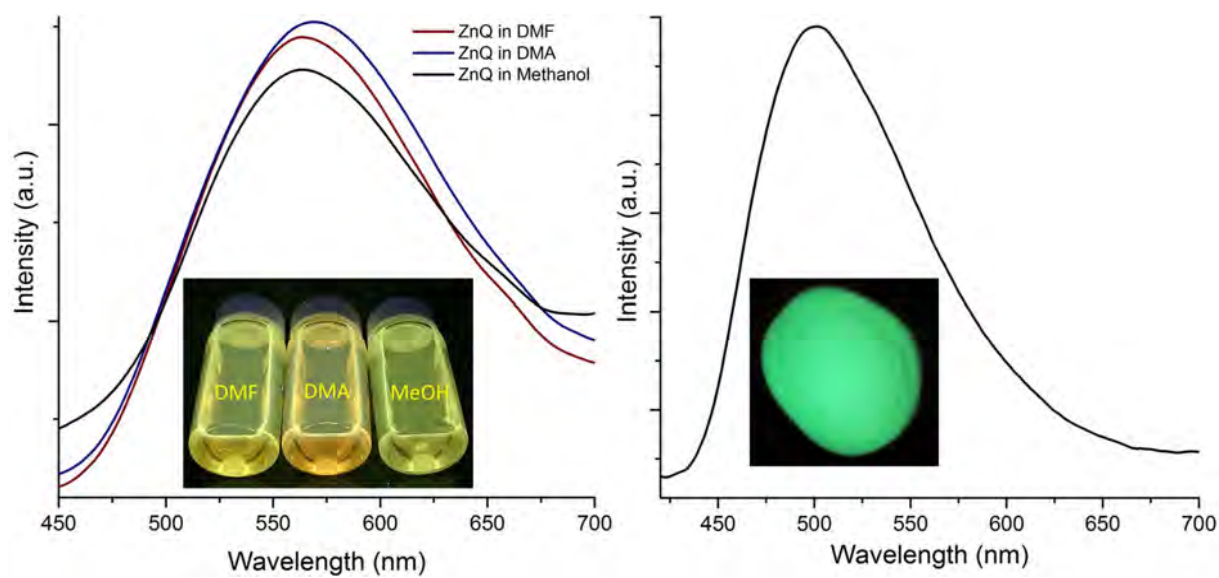

**Figure S28.** (a) Red-shifted emission spectra ($\lambda_{em} > 500$ nm) of pure ZnQ emitter complex dispersed in three different solvents: N,N-Dimethylformamide (DMF), N,N-Dimethylacetamide (DMA) and methanol. (b) Solid-state emission of pure ZnQ compound with $\lambda_{em} = 500$ nm. We studied emission behavior of the pure ZnQ complex in both liquid- and solid-states to achieve insights into the plausible molecular mechanism without nanoconfinement of MOF pores. In solid state (right), pure ZnQ emission is ~500 nm, however, when dispersed as solutions (left) in DMF, DMA or methanol, a red shift was observed at similar intensities. This result suggests intermolecular interaction is promoted in solution state due to its higher molecular mobility, originating from molecular overlapping to form strong aromatic π-π and H-bonding interactions, which are impeded in the solid state.

## 12 References

[1]     Burrows AD, Cassar K, Friend RMW, Mahon MF, Rigby SP, Warren JE. Solvent hydrolysis and templating effects in the synthesis of metal-organic frameworks, CrystEngComm 2005;7:548.

[2]     Biemmi E, Bein T, Stock N. Synthesis and characterization of a new metal organic framework structure with a 2D porous system: $(H_2NEt_2)_2[Zn_3(BDC)_4]\cdot3DEF$, Solid State Sci. 2006;8:363.

[3]     Macrae CF, Bruno IJ, Chisholm JA, Edgington PR, McCabe P, Pidcock E, Rodriguez-Monge L, Taylor R, van de Streek J, Wood PA. Mercury CSD 2.0 - new features for the visualization and investigation of crystal structures, J. Appl. Crystallogr. 2008;41:466.

[4]     Delley B. Time dependent density functional theory with DMol3, J Phys Condens Matter 2010;22:384208.

[5]     Patey MD, Dessent CEH. A PW91 Density Functional Study of Conformational Choice in 2-Phenylethanol,n-Butylbenzene, and Their Cations:  Problems for Density Functional Theory?, The Journal of Physical Chemistry A 2002;106:4623.

[6]     Inada Y, Orita H. Efficiency of numerical basis sets for predicting the binding energies of hydrogen bonded complexes: evidence of small basis set superposition error compared to Gaussian basis sets, J. Comput. Chem. 2008;29:225.

[7]     Chan MK, Ceder G. Efficient band gap prediction for solids, Phys. Rev. Lett. 2010;105:196403.

[8]     Kamlet MJ, Abboud JLM, Abraham MH, Taft RW. Linear Solvation Energy Relationships .23. A Comprehensive Collection Of The Solvatochromic Parameters, Pi-Star, Alpha And Beta, And Some Methods For Simplifying The Generalized Solvatochromic Equation, J. Org. Chem. 1983;48:2877.